\documentclass[]{article}
\usepackage[margin=25mm]{geometry}
\usepackage{authblk}
\usepackage[utf8]{inputenc}
\usepackage{graphicx}
\usepackage{dcolumn}
\usepackage{bm}
\usepackage{amsmath,amsfonts,amssymb}
\usepackage{commath}
\usepackage{tabularx}
\usepackage[caption=false]{subfig}
\usepackage{float}
\usepackage{threeparttable}
\usepackage{hyperref}
\hypersetup{colorlinks,allcolors=blue}
\usepackage{cleveref}

\title{Fractal dimension analysis of spatio-temporal patterns using image processing  and nonlinear time-series analysis}
\author[1]{Debasmita Banerjee}%
\author[1]{Amit Kumar Jha}
\author[1]{A.N.Sekar Iyengar}
\author[1]{M.S.Janaki}
\affil[1]{Plasma Physics Division, Saha Institute of Nuclear Physics, HBNI, 1/AF, Bidhannagar, Kolkata 700064, India}
\date{}
\begin{document}
\maketitle
\begin{abstract}

This article deals with the estimation of fractal dimension of spatio-temporal patterns that are generated by numerically solving the Swift Hohenberg (SH) equation. The patterns were converted into a spatial series (analogous to time series) which were shown to be chaotic by evaluating the largest Lyapunov exponent. We have applied several nonlinear time-series analysis techniques like Detrended fluctuation and Rescaled range on these spatial data to obtain Hurst exponent values that reveal spatial series data to be long range correlated. We have estimated fractal dimension from the Hurst and power law exponent and found the value lying between 1 and 2. The novelty of our approach lies in estimating fractal dimension using image to data conversion and spatial series analysis techniques, crucial for experimentally obtained images. 
\end{abstract}

\section{Introduction}
The ubiquitous presence of fractals have long motivated the scientific community to quantify nature's geometrical complexity. According to historical records, the term was first introduced by Benoit Mandelbrot in 1975 to expand the concept of theoretical fractional dimensions to geometric patterns in nature~\cite{Ref. 1}. A defining characteristic of fractals is its infinite number of scale lengths. Due to this, the structure of a fractal is identical in response to random magnification or scale variation. A mathematical characterizing quantity of fractals is called fractal dimension (FD) that measures complexity of patterns as a ratio of the change in structure to the change in scale. Our study involves estimation of the Fractal Dimension of the spatial data generated from the evolution patterns of Swift Hohenberg equation. This equation is a parabolic partial differential equation of fourth order known for modeling patterns in simple fluids (e.g. Rayleigh-Bénard convection) and in a diverse collection of complex fluids and biological formation. Previous theoretical measurements of fractal dimensions have been mainly obtained by box counting method~\cite{Ref. 2} which involves analyzing complex patterns by scanning and breaking dataset ( or images ) in finer pieces.
In our work, applying numerical solving techniques, we have obtained non-linear spatial series as solution of the SH equation. These non-linear spatial series are then converted into color plots that give patterned images.

We have further retrieved the same non-linear spatial series by converting the image intensity into a spatial series (analogous to  time  series)  using  image processing  method  and  then performed  non-linear  spatial  series  analysis. The purpose was to demonstrate the usefulness of the method in case of experimental systems when the governing equation is not known and one only has images of the pattern but want to obtain non-linear spatial series for fractal dimension calculation.
\\
Analysis of the fractal dimension of the spatial series from the power law exponents reveals a non-integer value between 1 to 2 for our data. The merit of this alternative method also lies in its compatibility with other nonlinear time series analysis techniques such as Recurrence analysis etc. Having extensive applications on experimental images from varied sections of sciences including evolution biology and surface physics, makes it specifically appealing. The system shows long range correlation as the Hurst exponent generated by both the Detrended Fluctuation and Rescaled Range Analysis follows the range between 0.51 and 0.87. We have also calculated fractal dimension from the Hurst exponents which provides a comparable result with that of Hurst exponents, estimated from power law coefficient.

\begin{figure*}[h]
\centering
\subfloat[Image 1]{
  \includegraphics[width=45mm,scale=0.3]{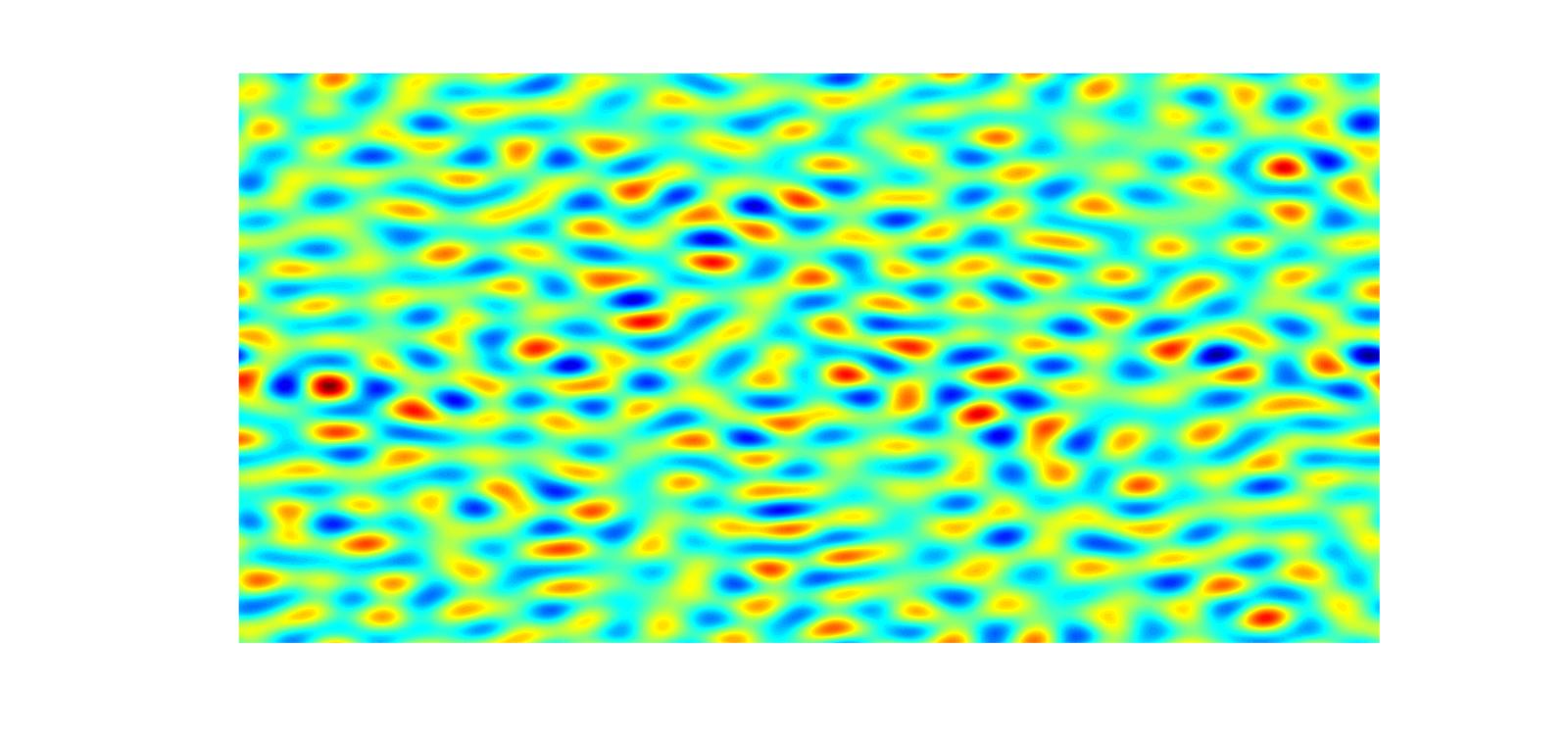}}
  \hspace*{-0.9em}
\subfloat[Image 2]{
  \includegraphics[width=45mm,scale=0.3]{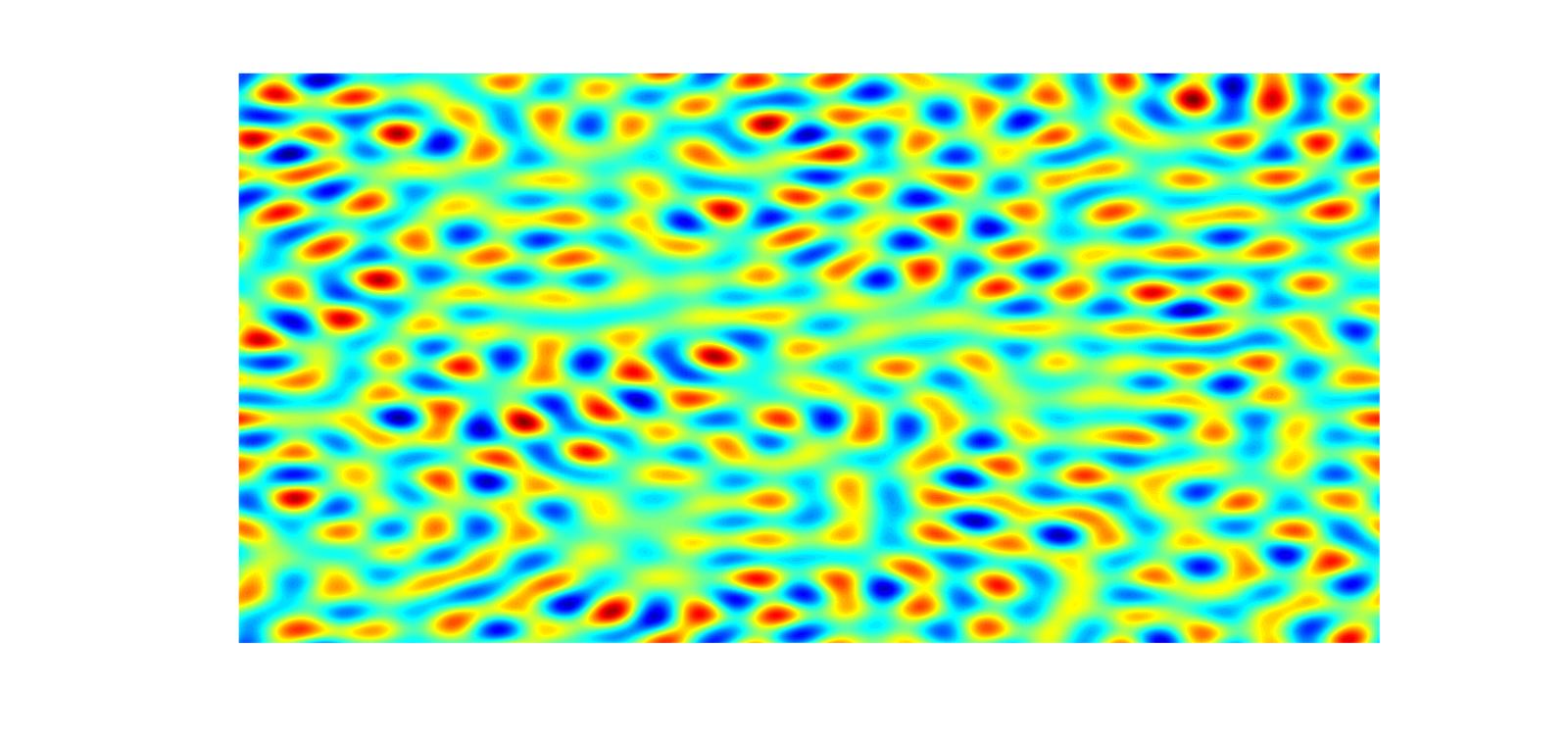}}
  \hspace*{-0.9em}
\subfloat[Image 3]{
 \includegraphics[width=45mm,scale=0.3]{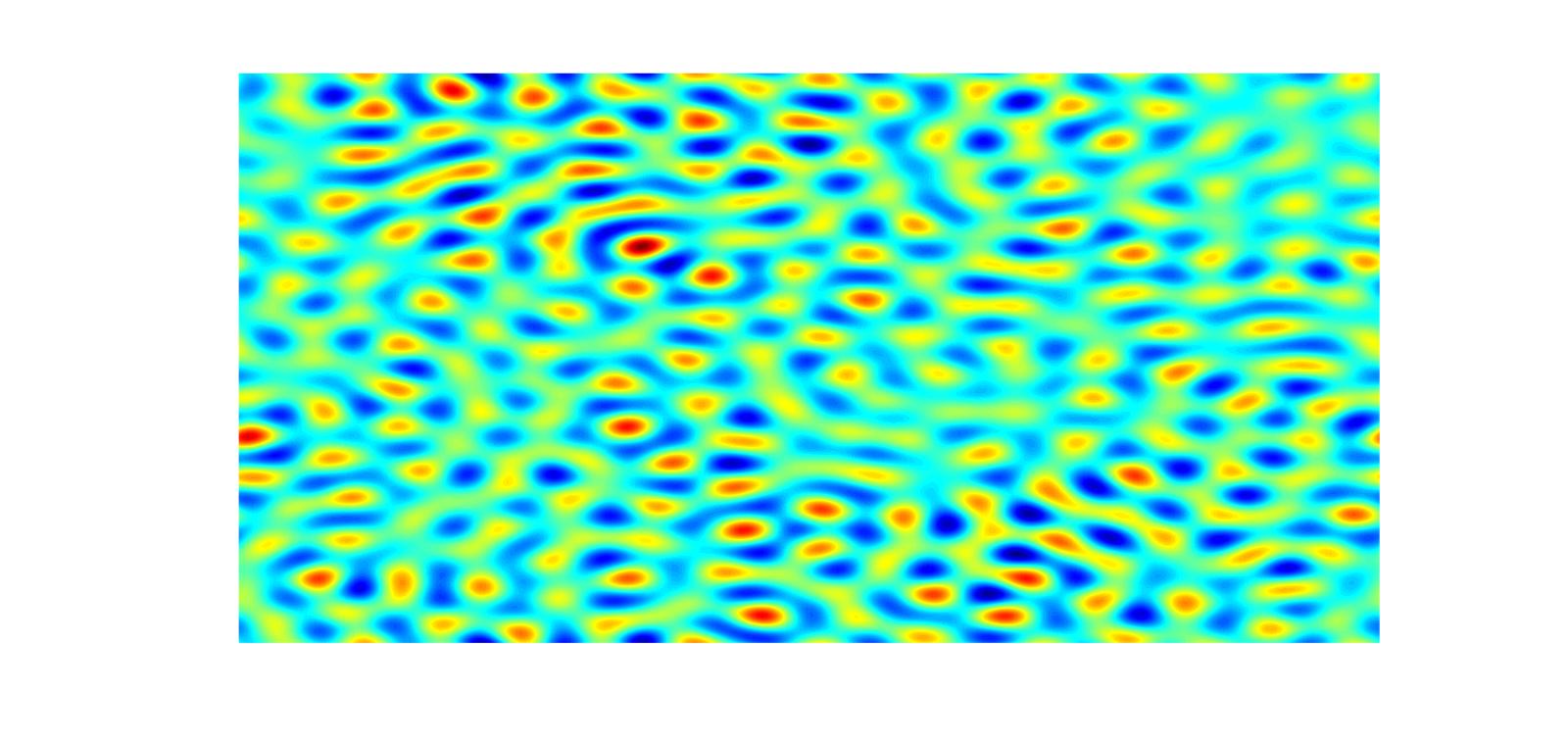}}
  \hspace*{-0.9em}
\subfloat[Image 4]{
  \includegraphics[width=45mm,scale=0.3]{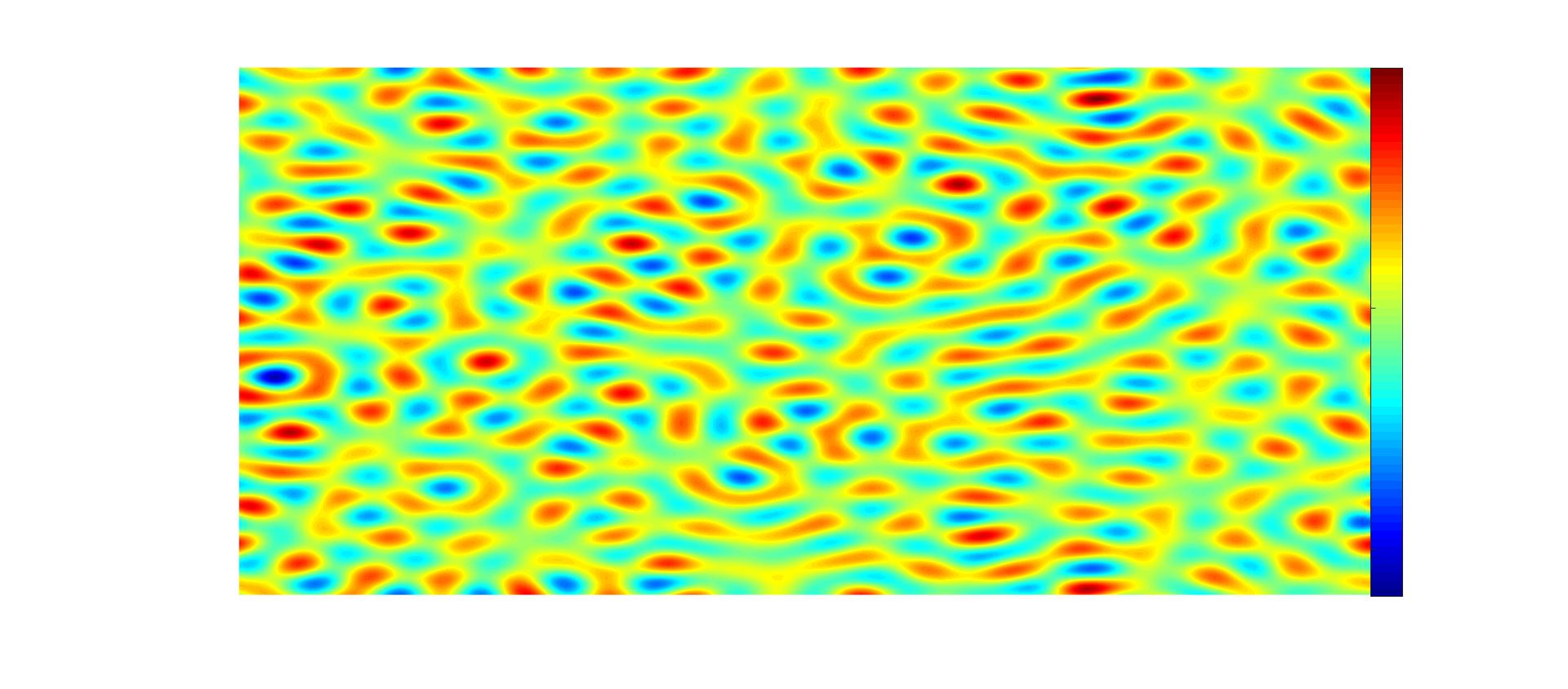}}
  \hfill
\subfloat[Image 5]{
  \includegraphics[width=45mm,scale=0.3]{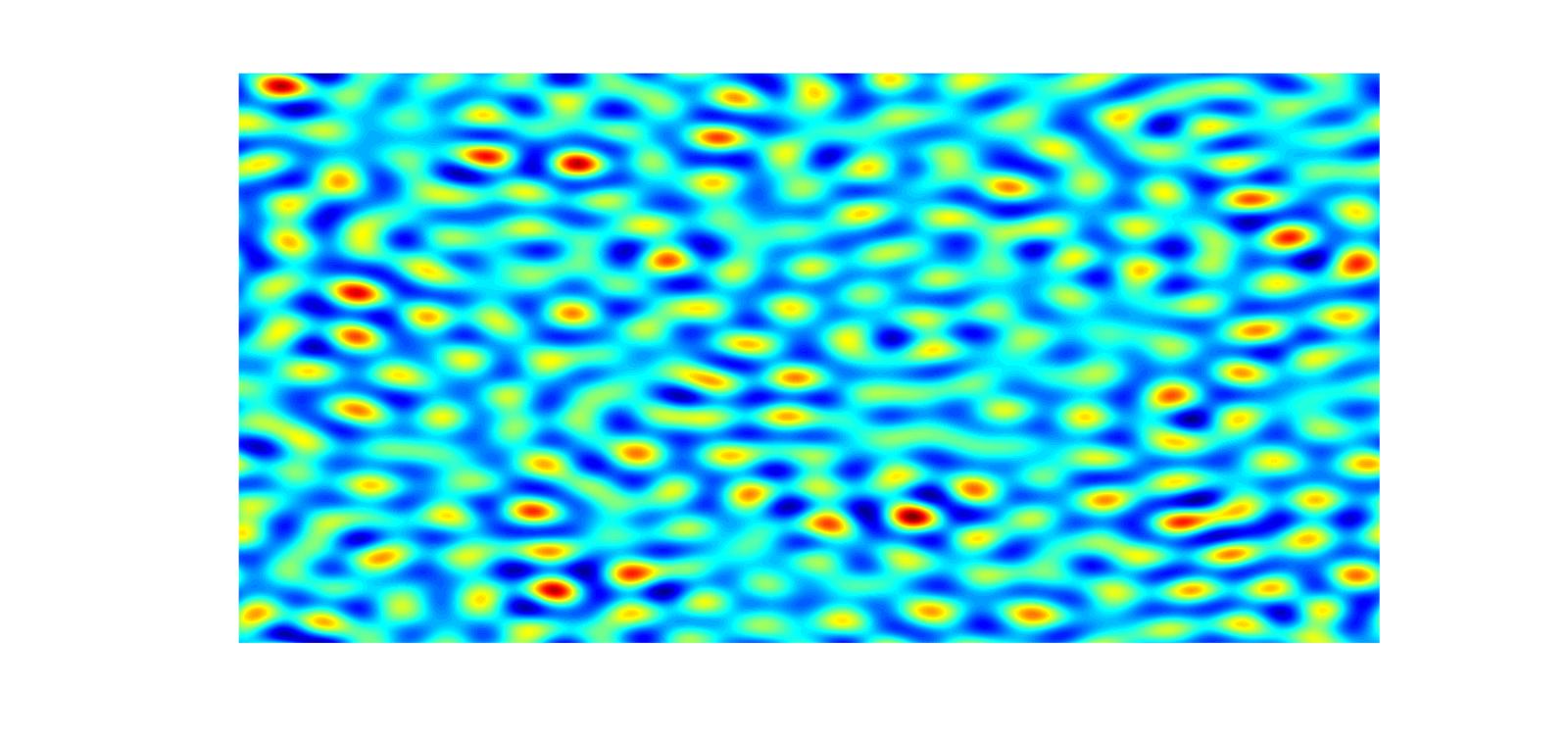}}
  \hspace*{-0.9em}
\subfloat[Image 6]{
  \includegraphics[width=45mm,scale=0.3]{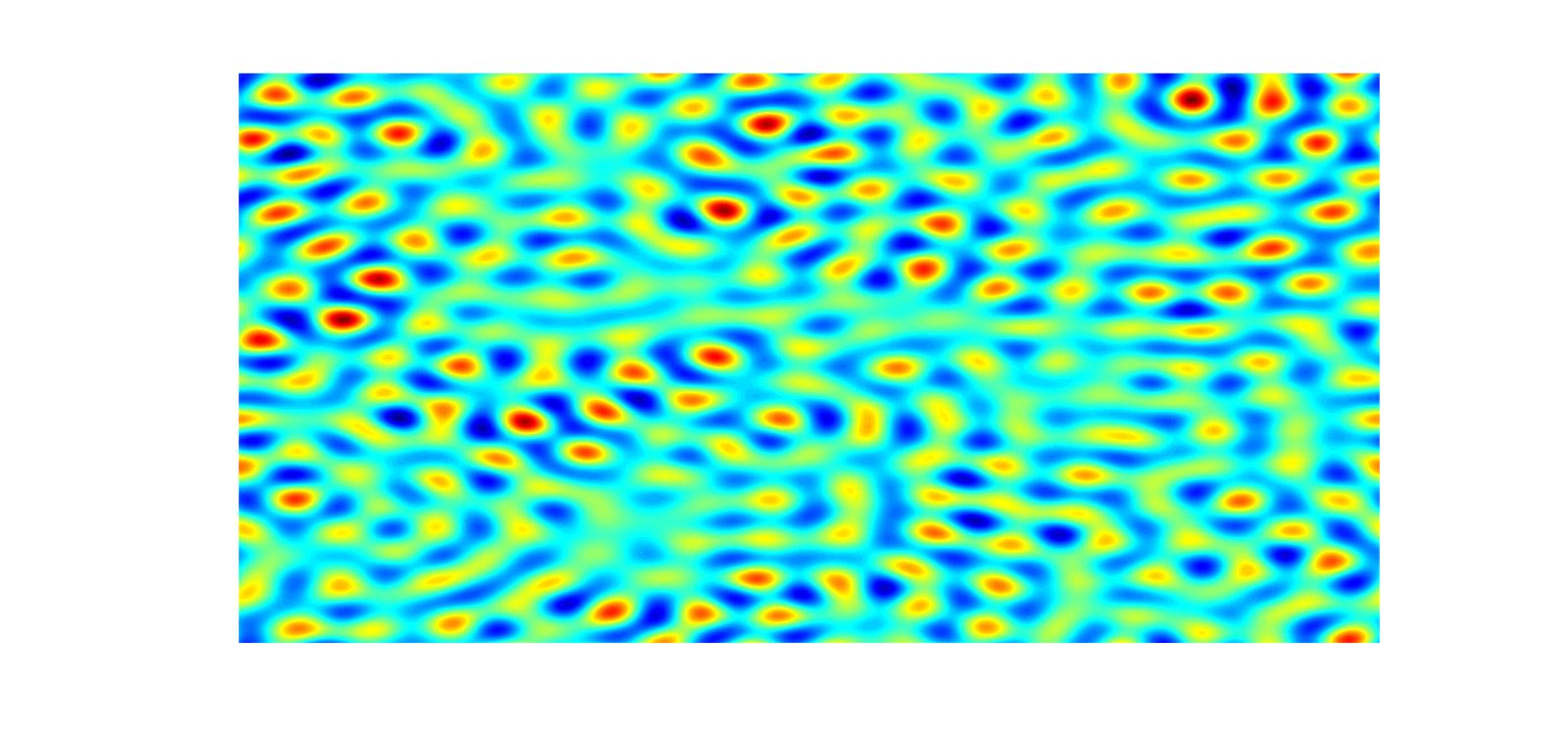}}
  \hspace*{-0.9em}
\subfloat[Image 7]{
  \includegraphics[width=45mm,scale=0.3]{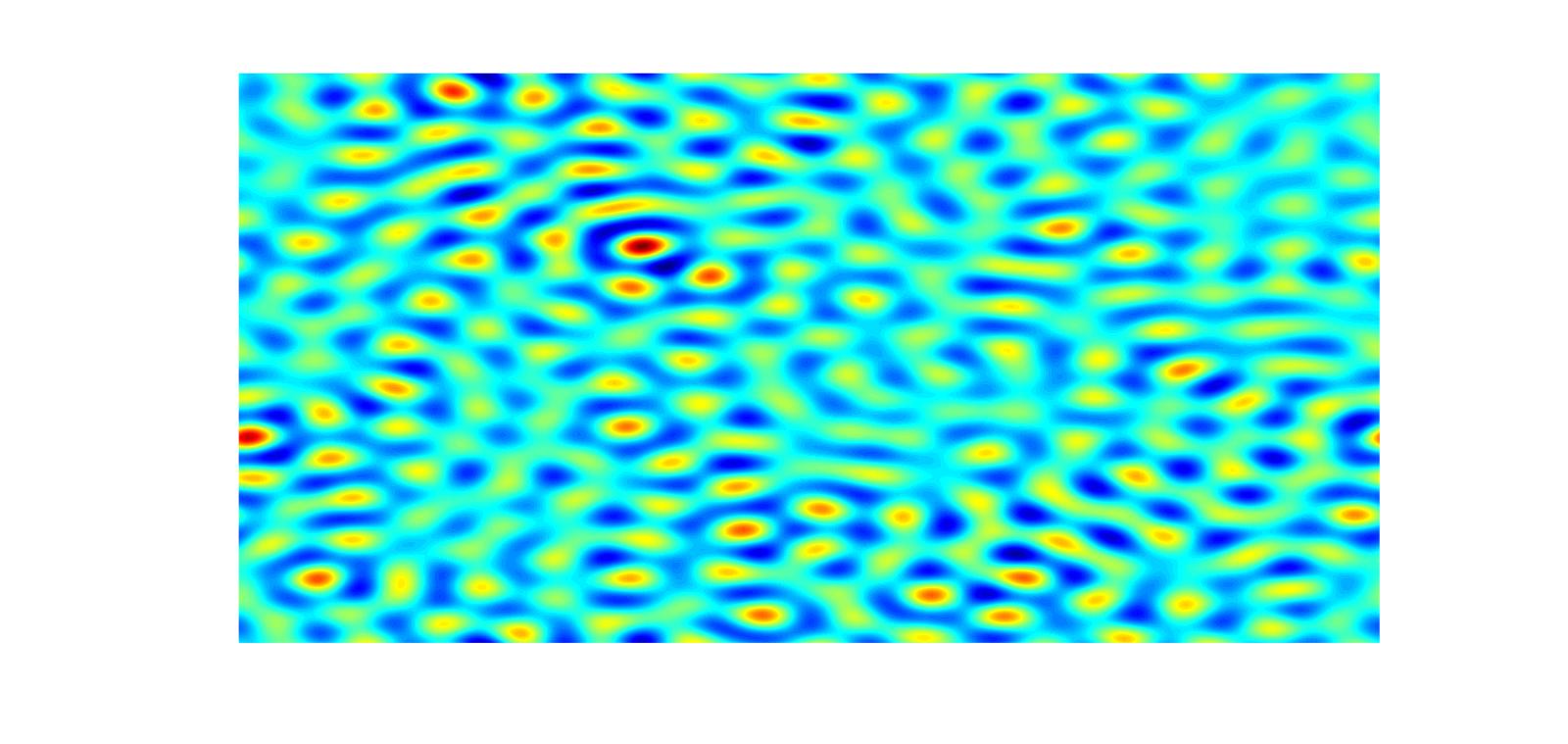}}
 \hspace*{-0.9em}
\subfloat[Image 8]{
  \includegraphics[width=45mm,scale=0.3]{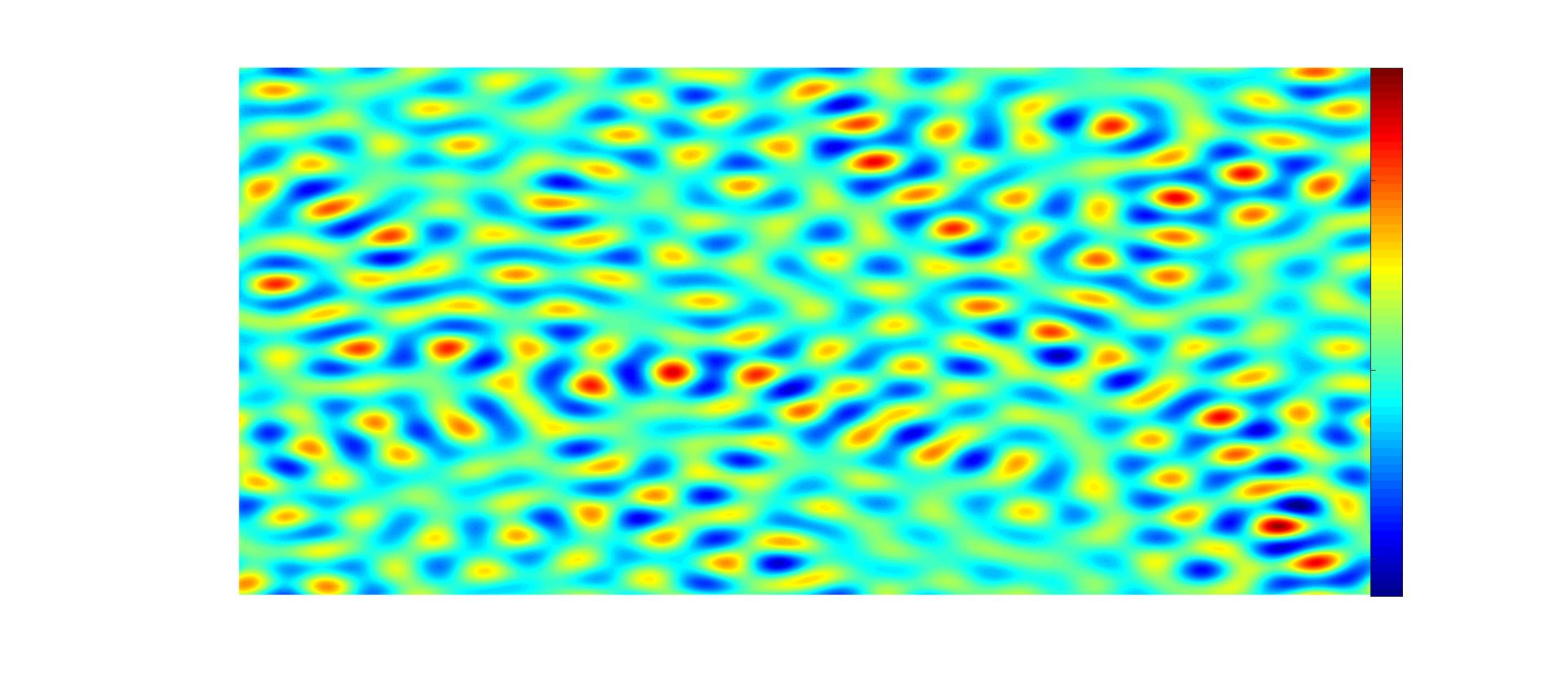}}
  \hfill
\subfloat[Image 9]{
  \includegraphics[width=45mm,scale=0.3]{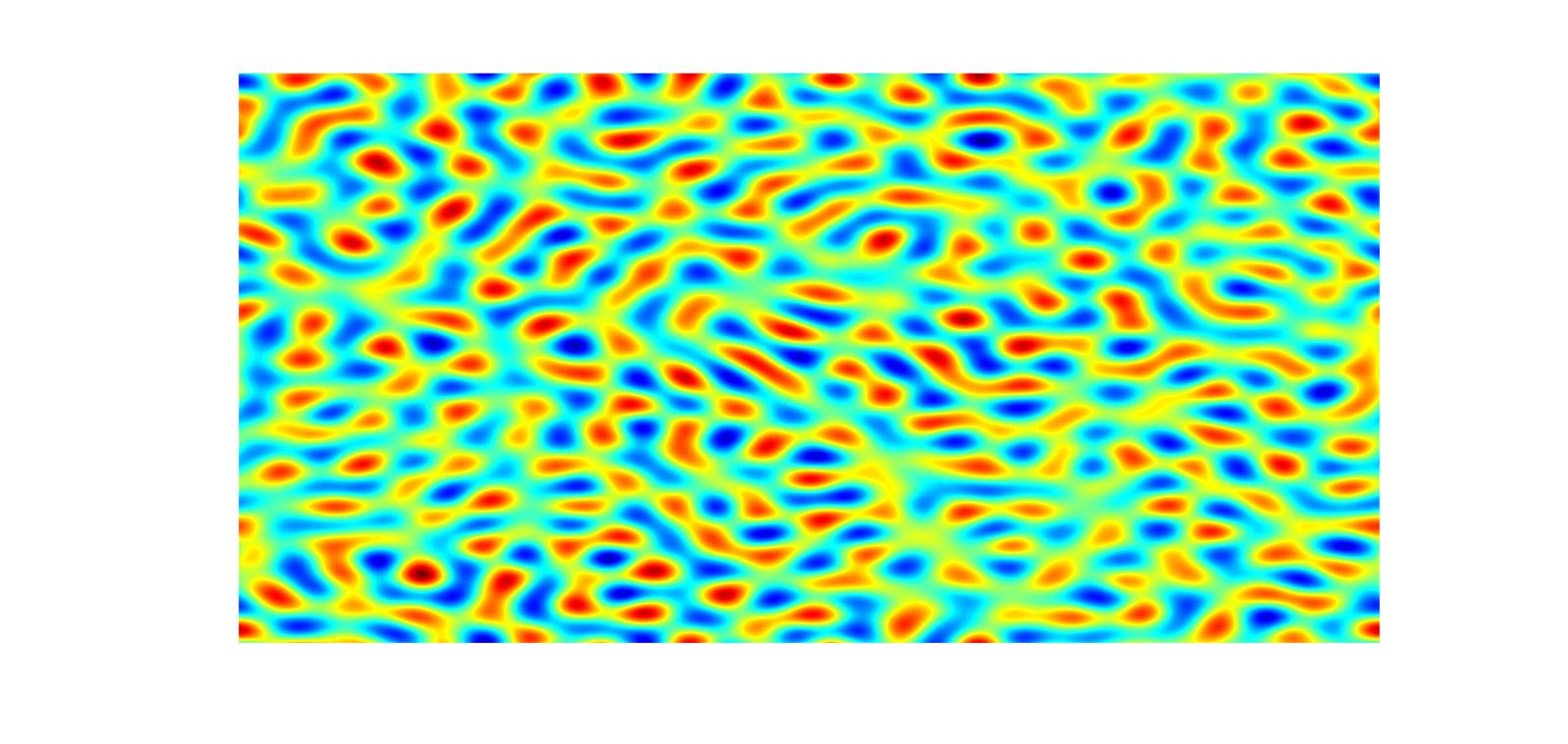}}
  \hspace*{-0.9em}
\subfloat[Image 10]{
  \includegraphics[width=45mm,scale=0.3]{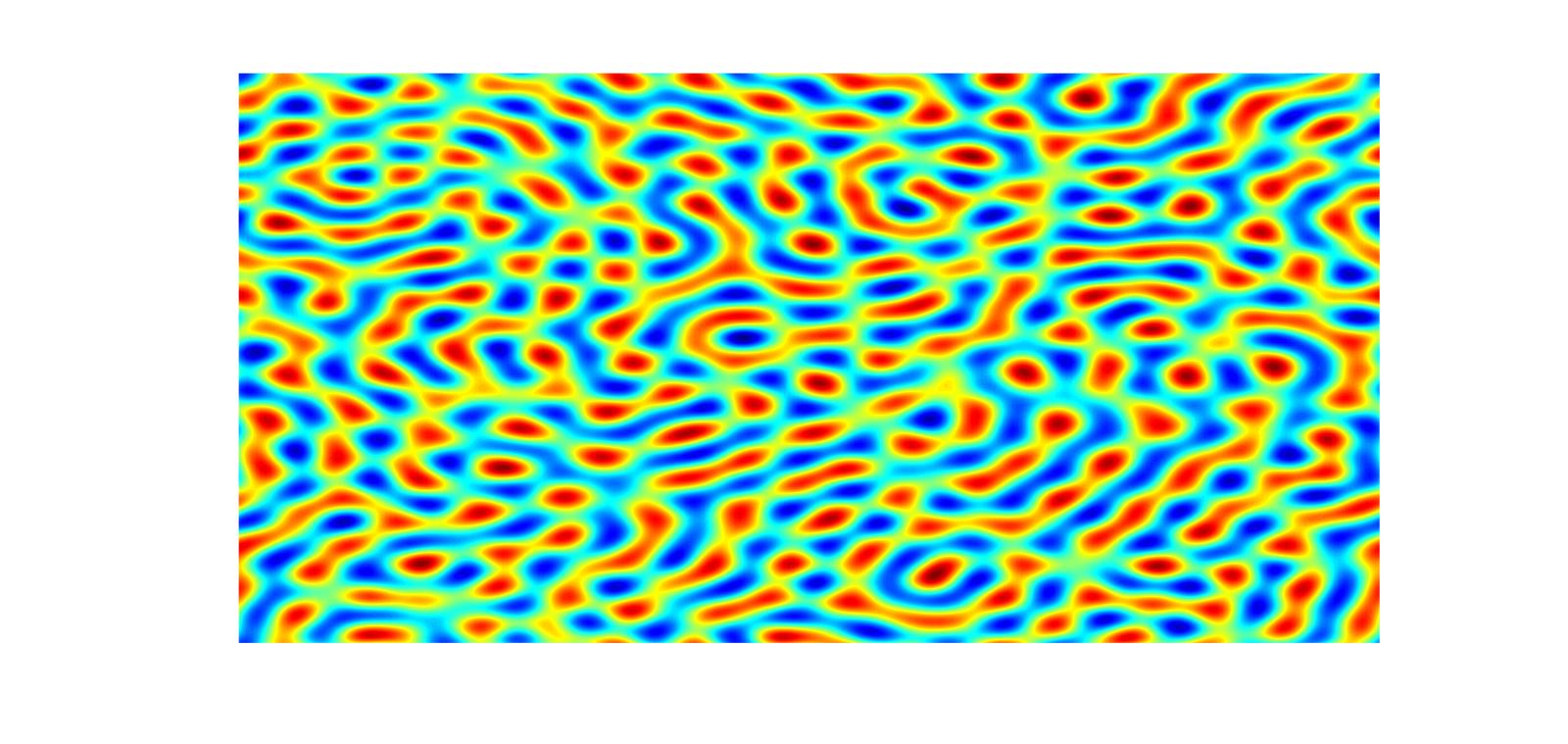}}
  \hspace*{-0.9em}
\subfloat[Image 11]{
  \includegraphics[width=45mm,scale=0.3]{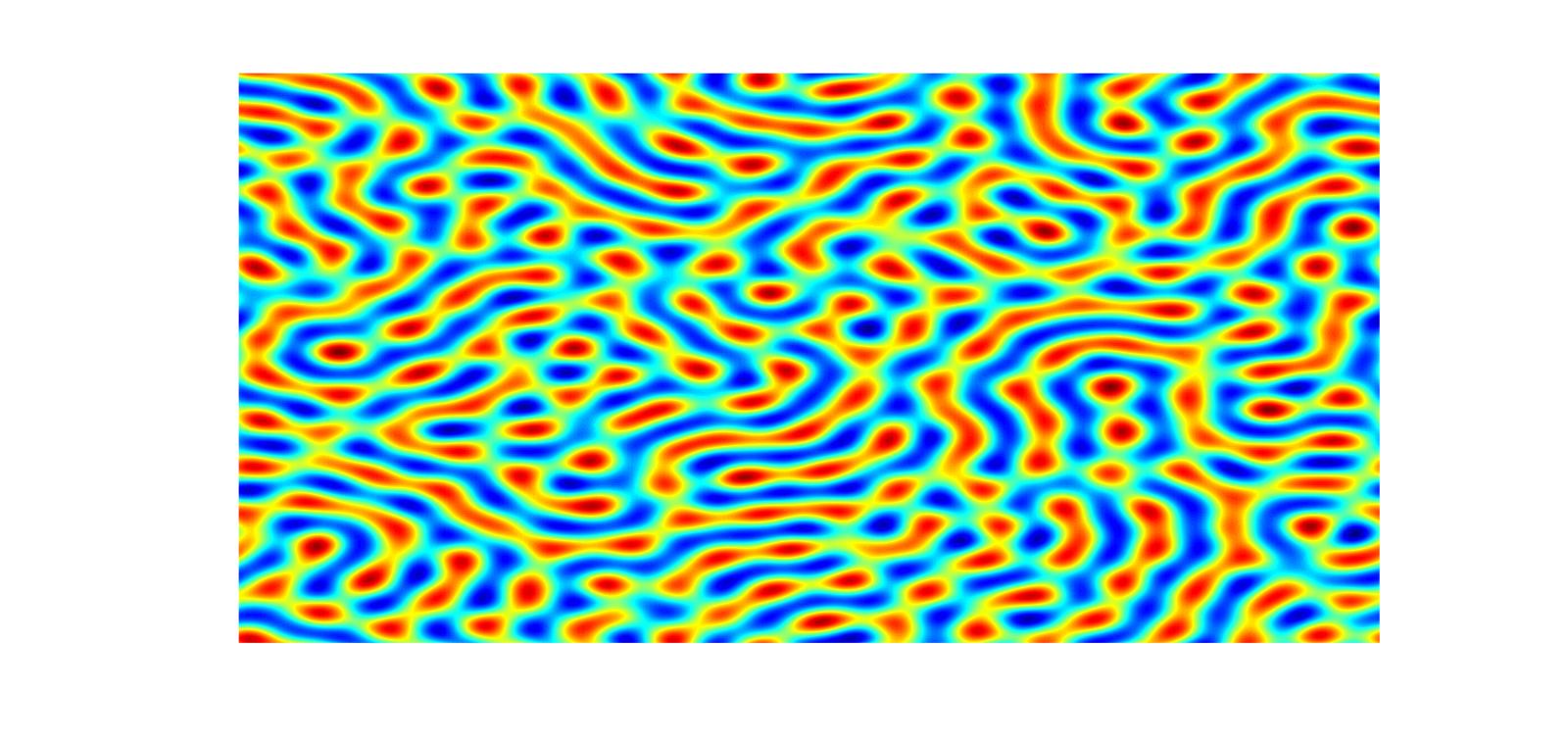}}
 \hspace*{-0.9em}
\subfloat[Image 12]{
  \includegraphics[width=45mm,scale=0.3]{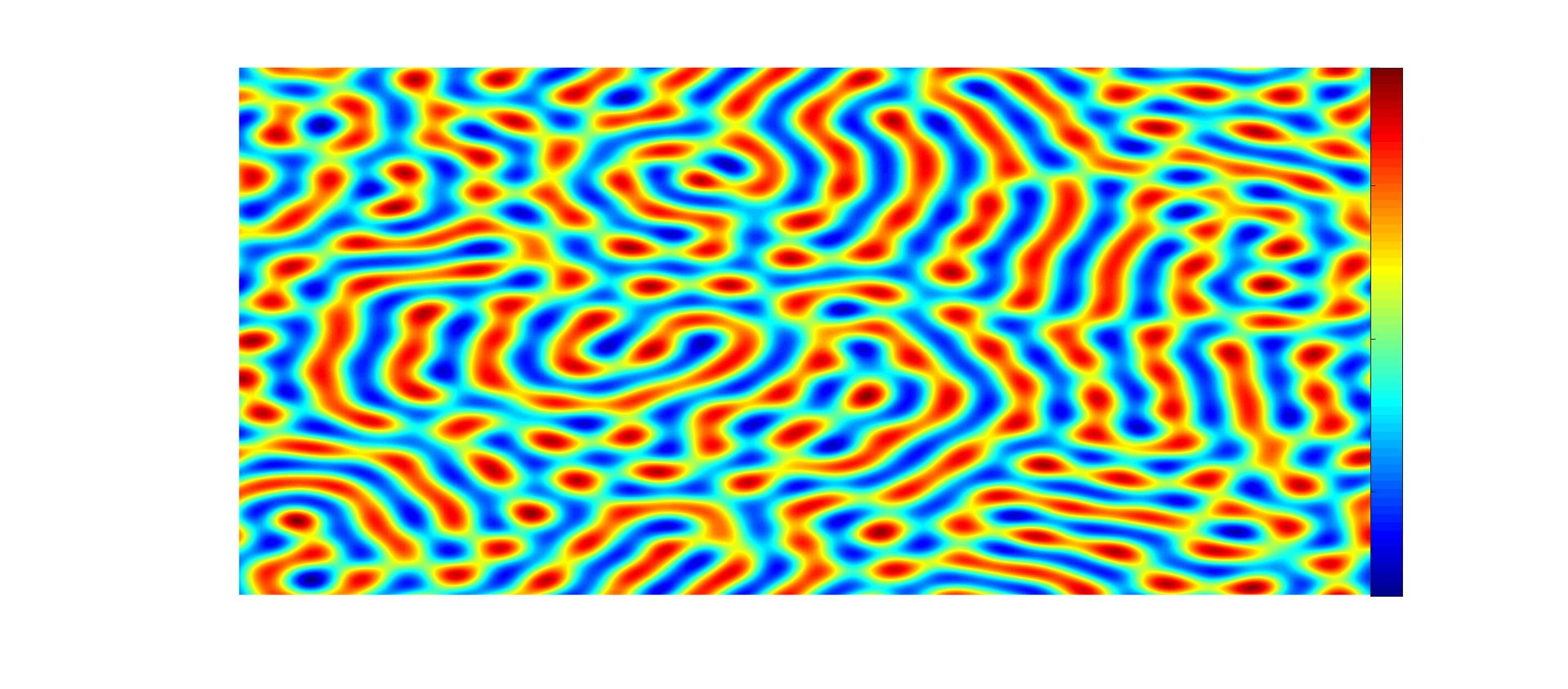}}
  \hfill
\subfloat[Image 13]{
  \includegraphics[width=45mm,scale=0.3]{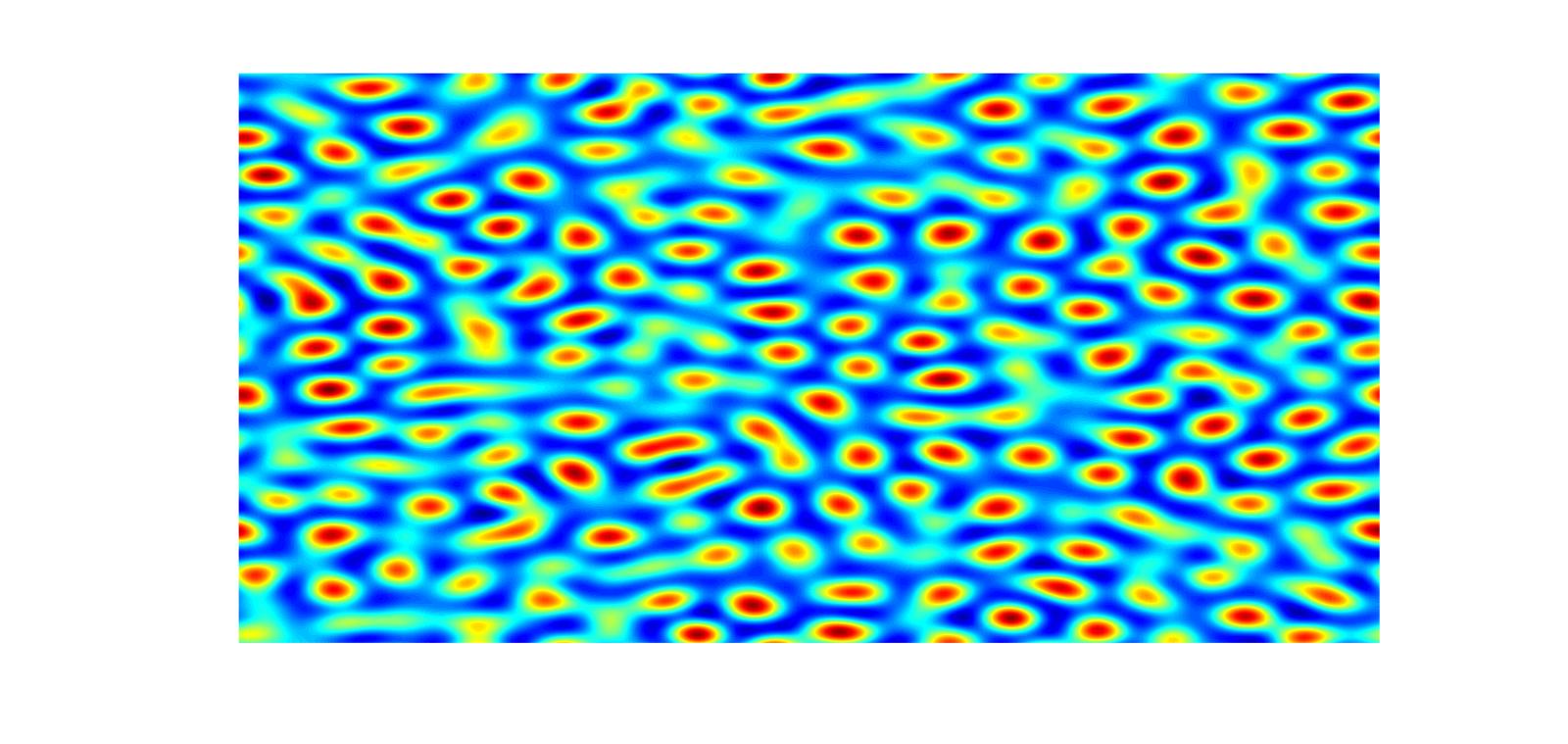}}
  \hspace*{-0.9em}
\subfloat[Image 14]{
  \includegraphics[width=45mm,scale=0.3]{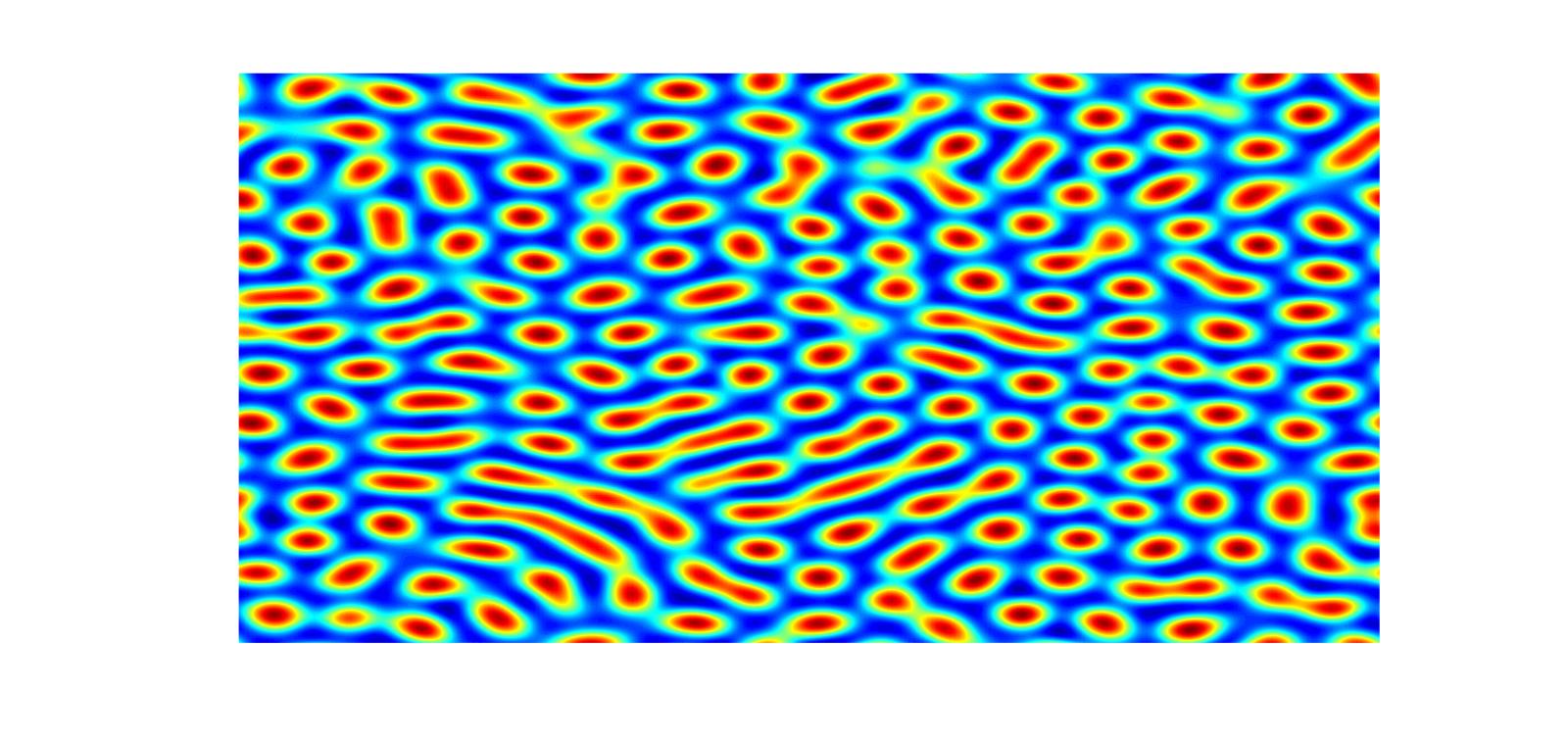}}
  \hspace*{-0.9em}
\subfloat[Image 15]{
  \includegraphics[width=45mm,scale=0.3]{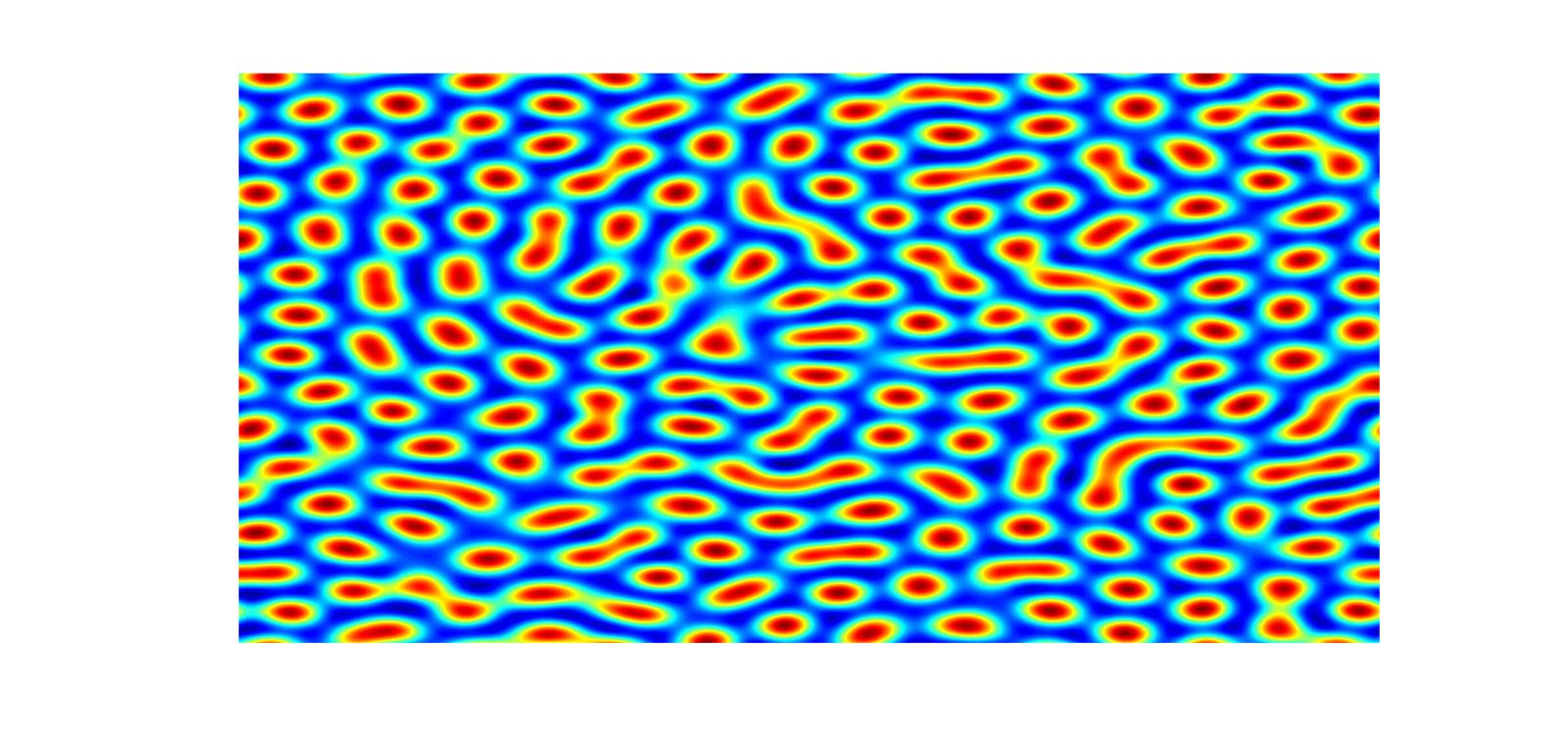}}
  \hspace*{-0.9em}
\subfloat[Image 16]{
  \includegraphics[width=45mm,scale=0.3]{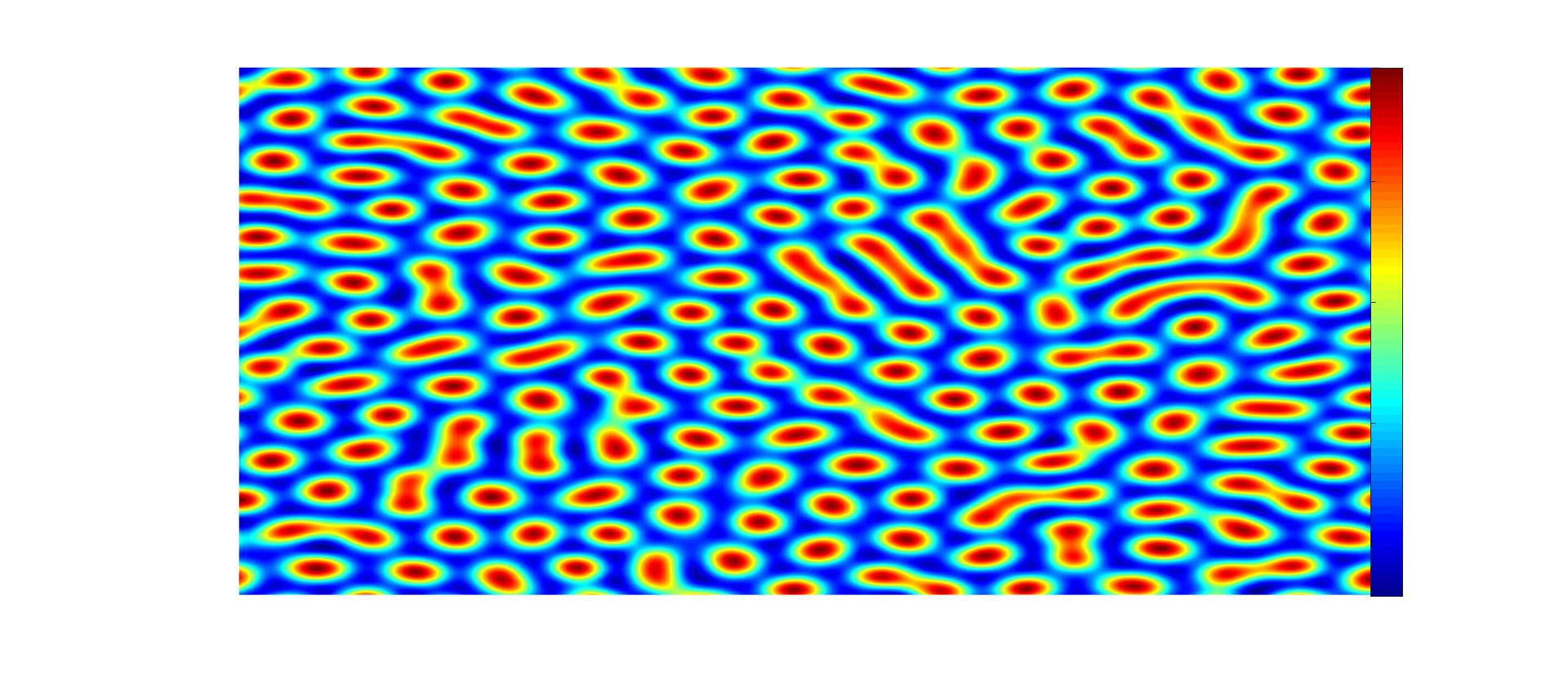}}
  \hfill
   
\caption{SH-solution producing a spatial evolution over “time” exposure for (a)Image 1 at $v = 0.1, \epsilon = 0.3, t = 5$ seconds (b) Image 2 at $ v= 0.1, \epsilon  = 0.3, t = 10 $ seconds (c) Image 3 at $v = 0.1, \epsilon = 0.3, t = 15$ (d) Image 4 (with colorbar) at $v = 0.1, \epsilon = 0.3, t = 20$ seconds (e) Image 5 at $v = 1.0, \epsilon = 0.3, t = 5 $ seconds (f) Image 6 at $v = 1.0, \epsilon = 0.3, t = 10 $ seconds (g) Image 7 at $v = 1.0, \epsilon = 0.3, t = 15 $ seconds (h) Image 8 (with colorbar) at $v = 1.0, \epsilon = 0.3, t = 20 $ seconds. (i)Image 9 at $v = 0.1, \epsilon = -0.3, t = 5$ seconds (j) Image 10 at $v = 0.1, \epsilon  = -0.3, t = 10$ seconds (k) Image 11 at $v = 0.1, \epsilon = -0.3, t = 15$ seconds  (l) Image 12 (with colorbar) at $v = 0.1, \epsilon = -0.3, t = 20$ seconds (m) Image 13 at $v = 1.0, \epsilon = -0.3, t = 5 $ seconds (n) Image 14 at $v = 1.0, \epsilon = -0.3, t = 10 $ seconds (o) Image 15  at $v = 1.0, \epsilon = -0.3, t = 15 $ seconds(p) Image 16 (with colorbar) at $v = 1.0, \epsilon = -0.3, t = 20 $ seconds}
\end{figure*}

The presence of chaos has recently been reported in the lab-grown experiment to study pattern in two dimensions~\cite{Ref. 3,Ref. 4} where the authors were able to detect system transition from long range correlation to anti-correlation between the initial and final state of the experimental patterns using a novel image processing technique.
Our current study is closely based on the proposed image to data conversion technique of above mentioned work.
 \begin{figure*}
 \centering
\subfloat[]{%
 \includegraphics[width=70mm,scale=0.3]{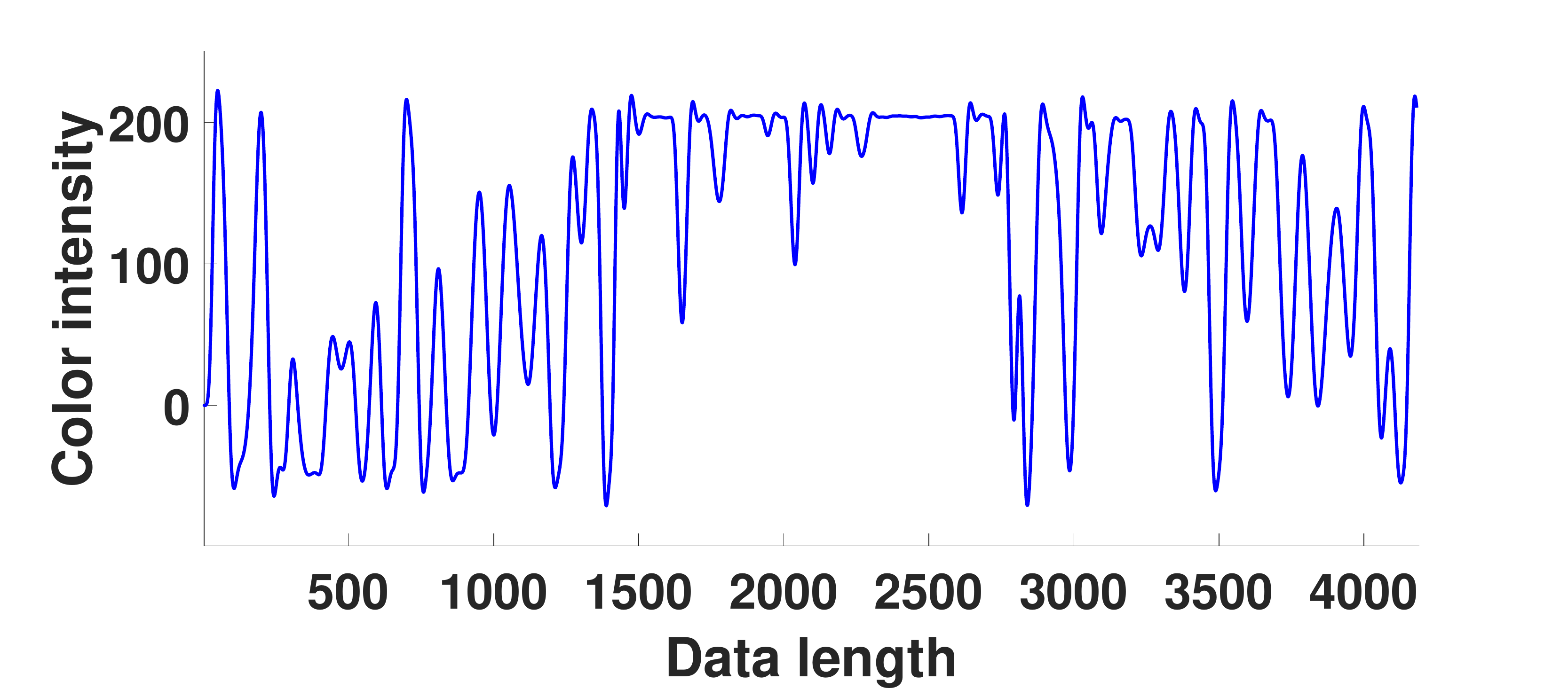}
}
\subfloat[]{%
  \includegraphics[width=70mm,scale=0.3]{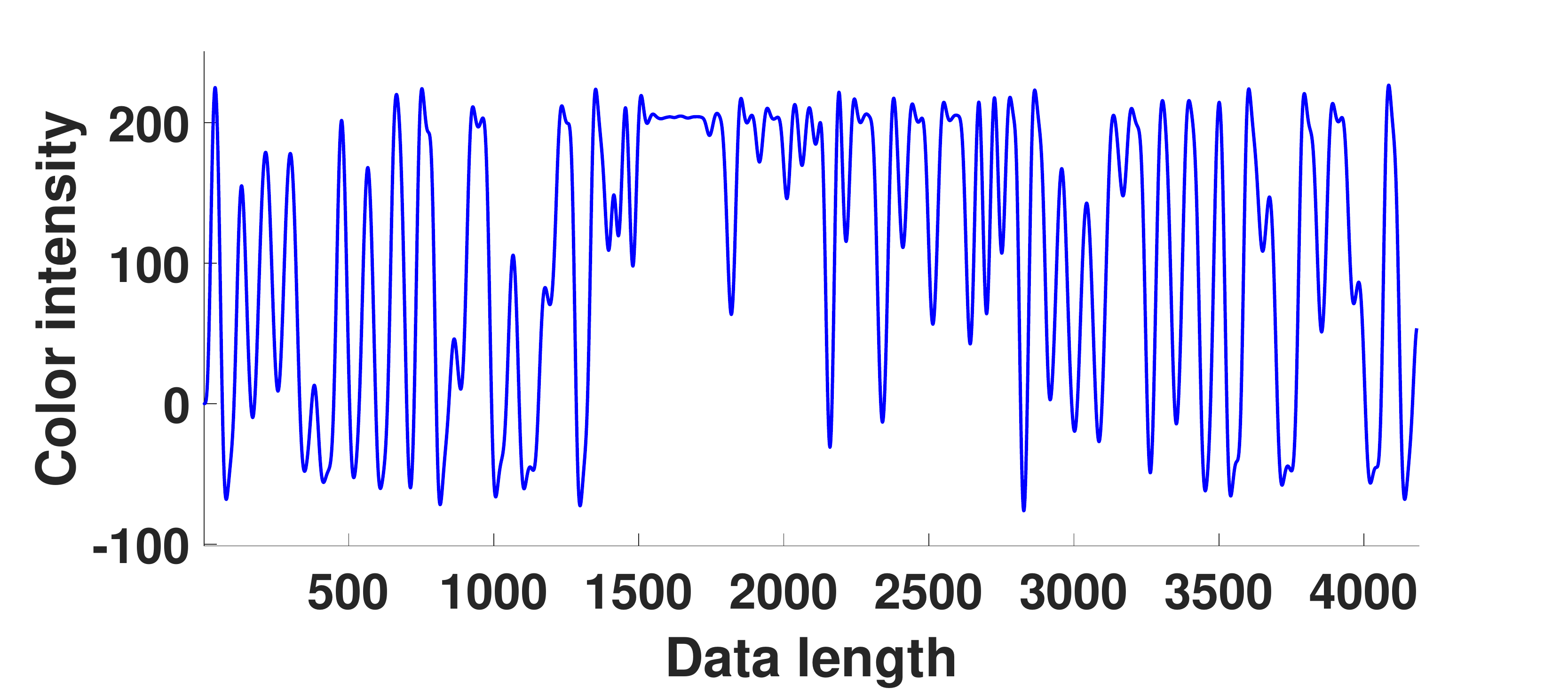}%
}\hfill
\subfloat[]{%
  \includegraphics[width=70mm,scale=0.3]{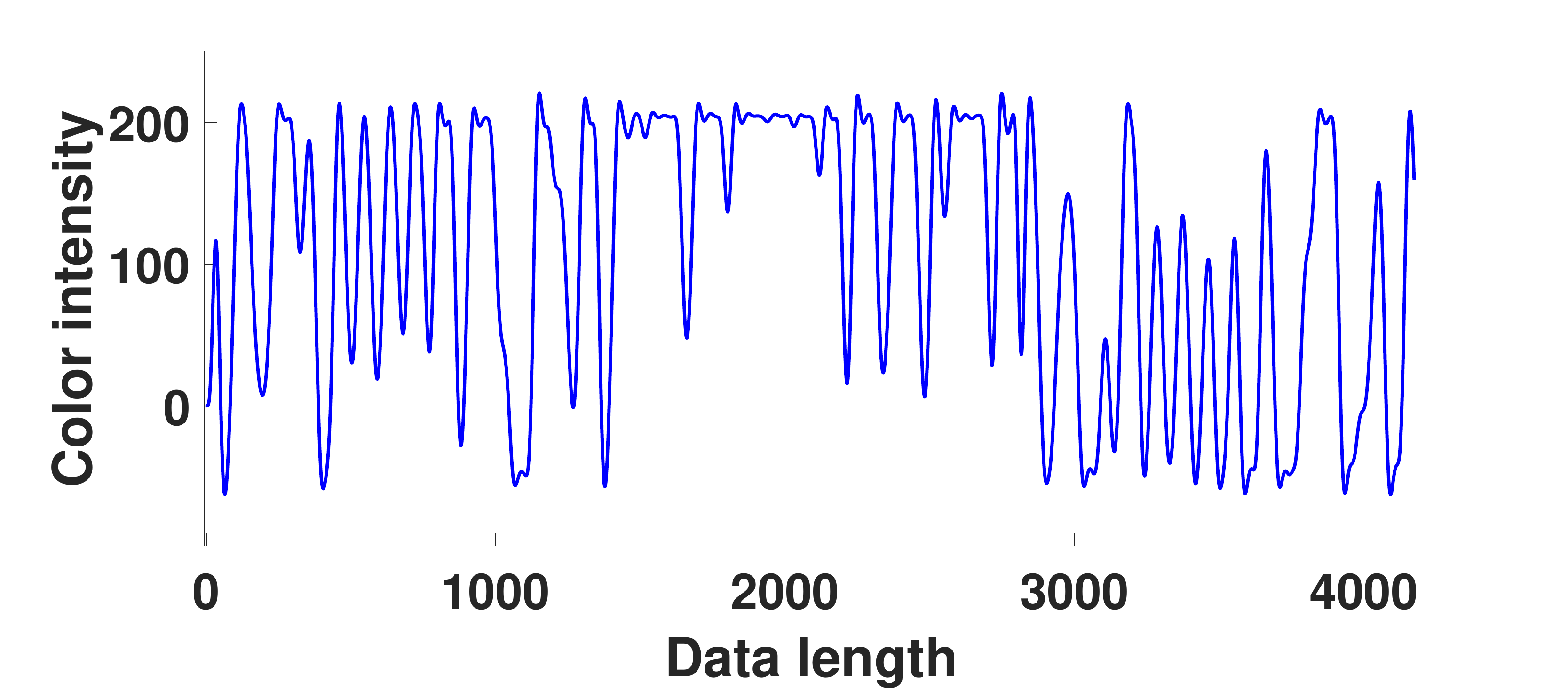}%
}
\subfloat[]{%
  \includegraphics[width=70mm,scale=0.3]{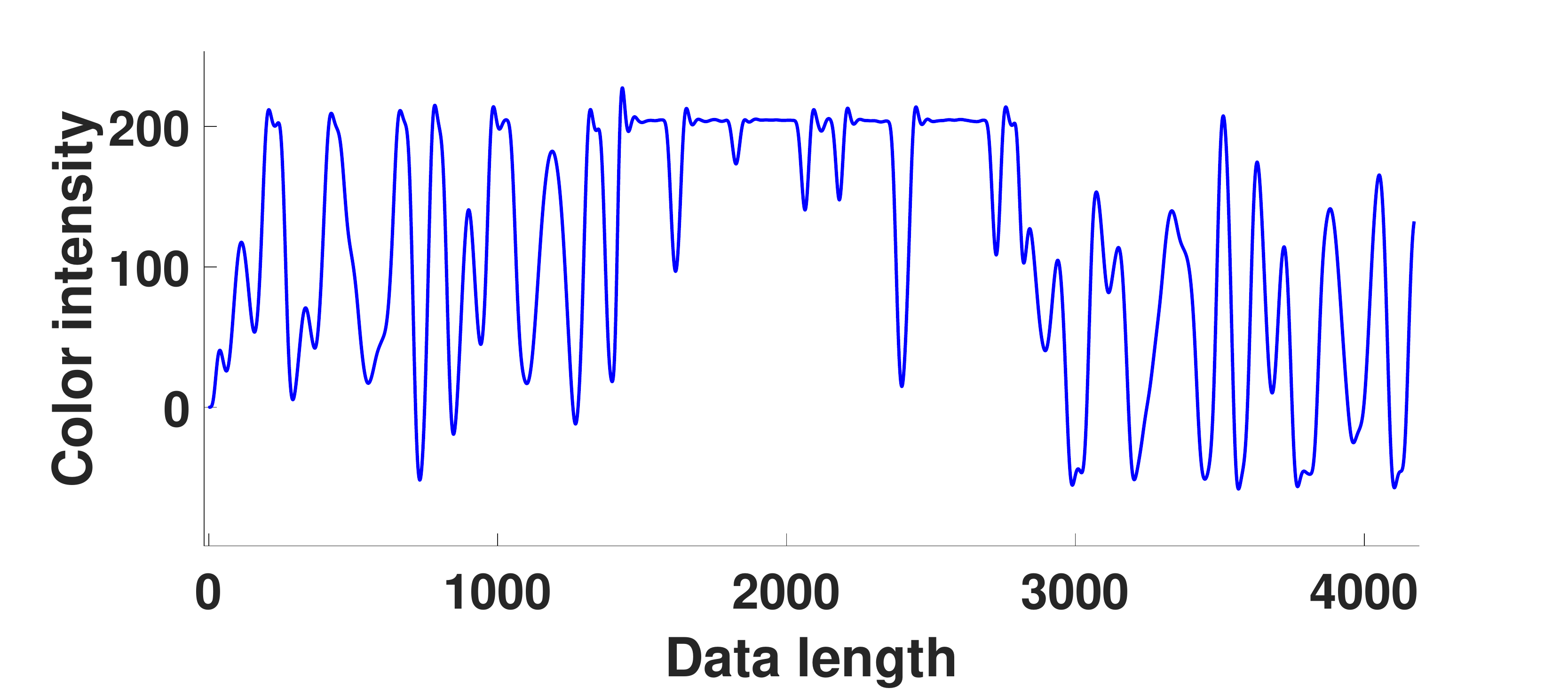}%
}\hfill
\subfloat{%
  \includegraphics[width=70mm,scale=0.3]{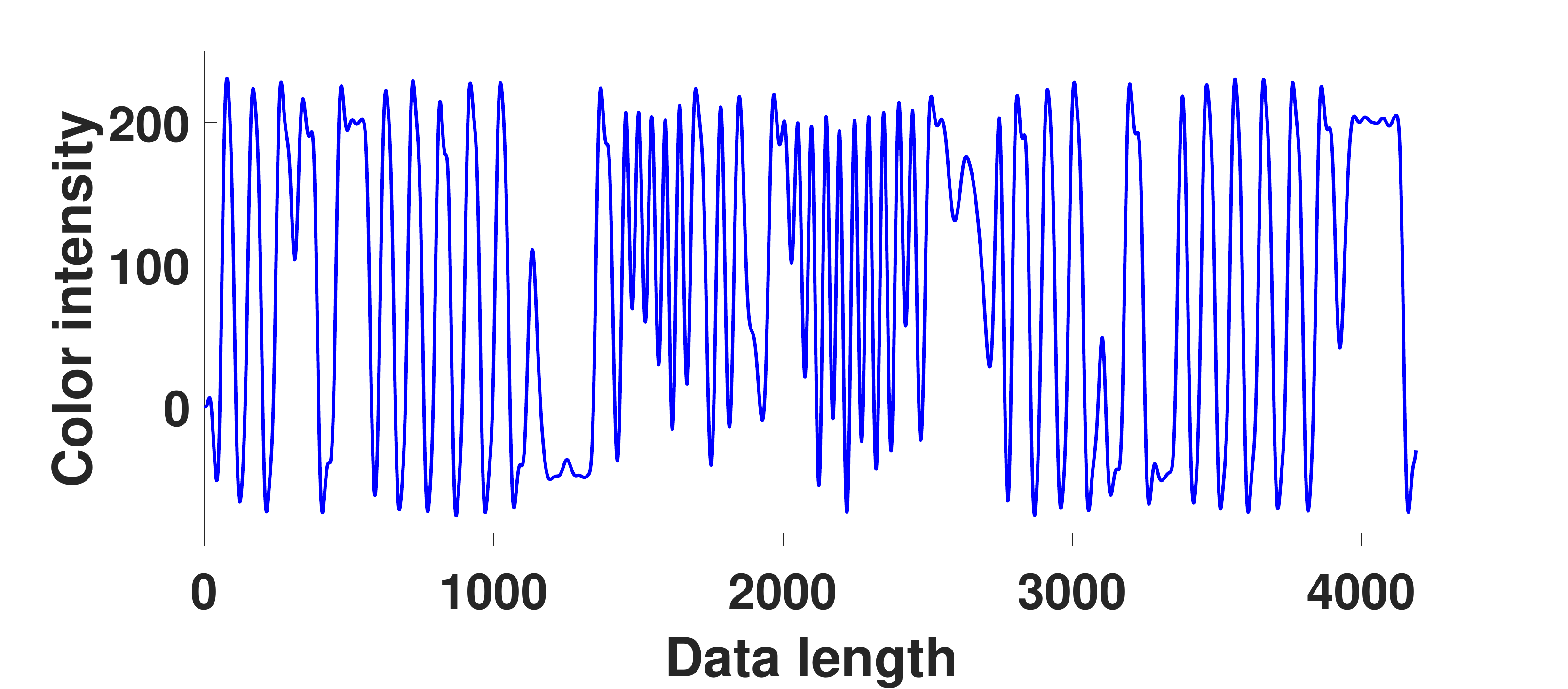}%
}
\subfloat{%
  \includegraphics[width=70mm,scale=0.3]{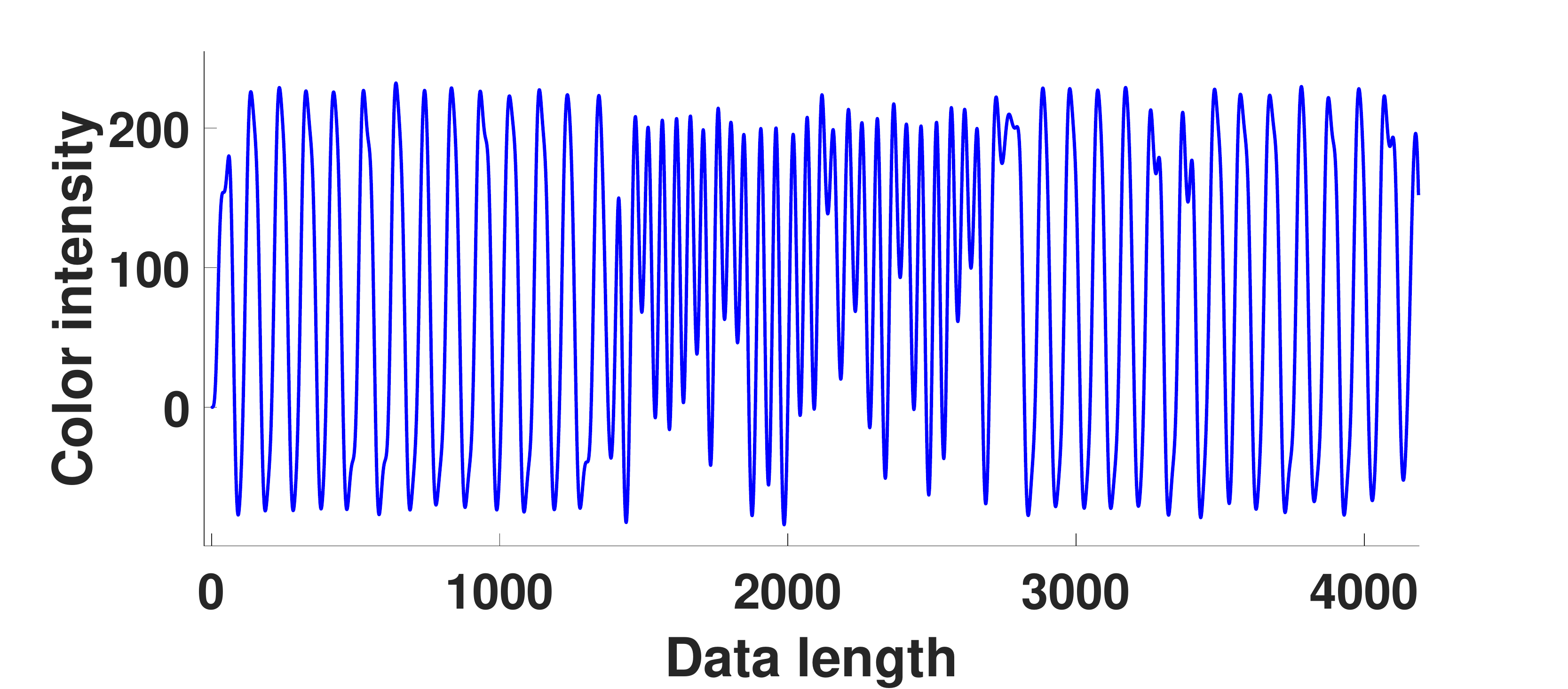}%
}\hfill

\caption{(a),(b) are generated from the 50th and 350th rows of Image 1,(c),(d) are generated from the spatial series of Image 4.(g),(h) are generated from the 50th and 350th rows of Image 12.}
\end{figure*}

\begin{figure*}
\subfloat[]{%
  \includegraphics[width=60mm,scale=0.3]{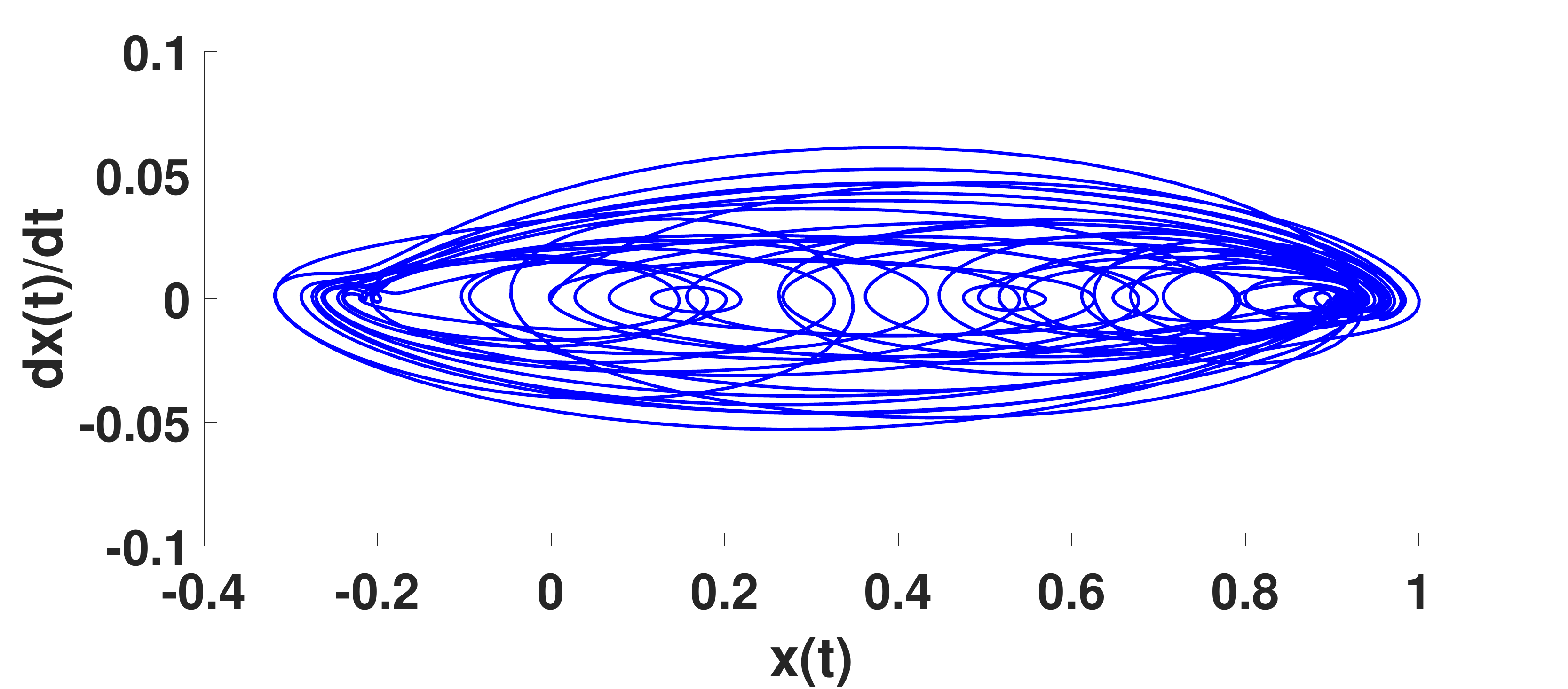}%
}
\subfloat[]{%
  \includegraphics[width=60mm,scale=0.3]{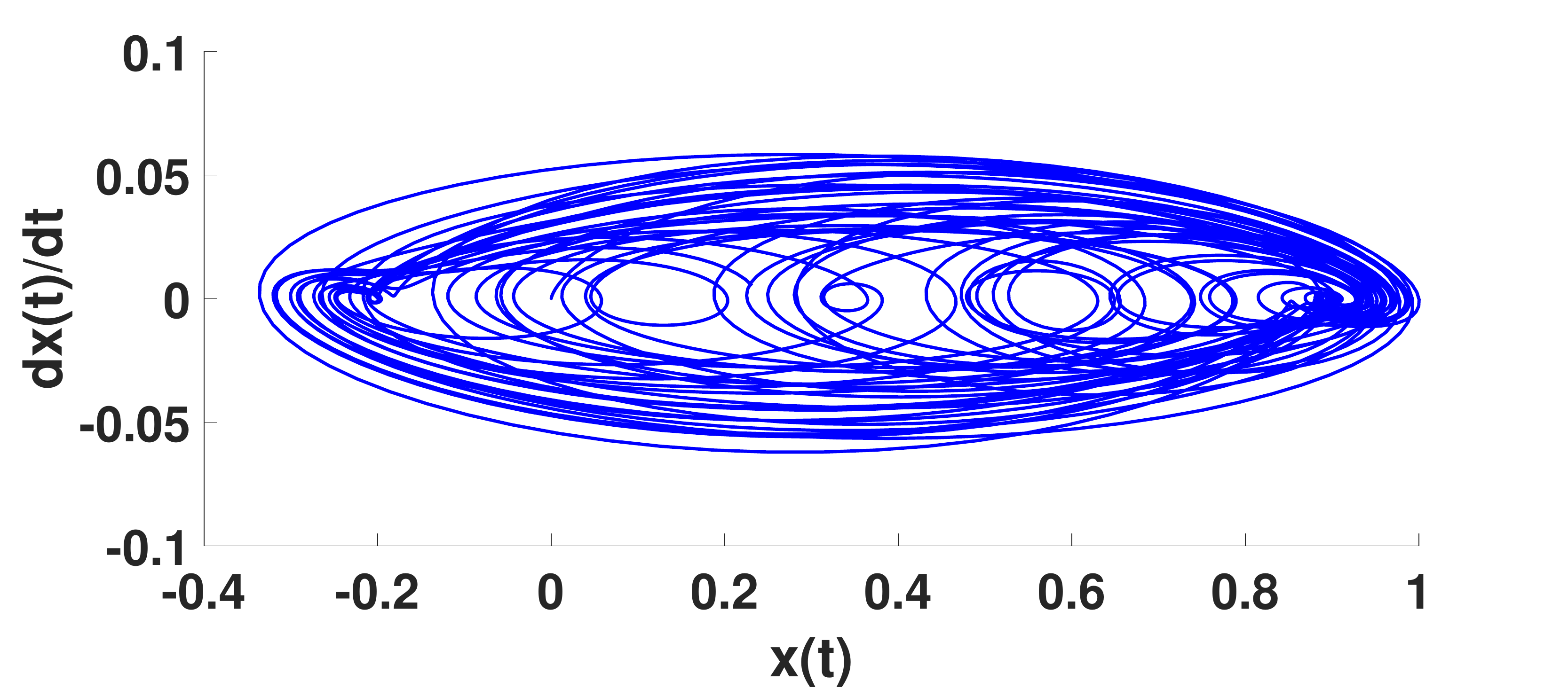}%
}
\subfloat[]{%
  \includegraphics[width=60mm,scale=0.3]{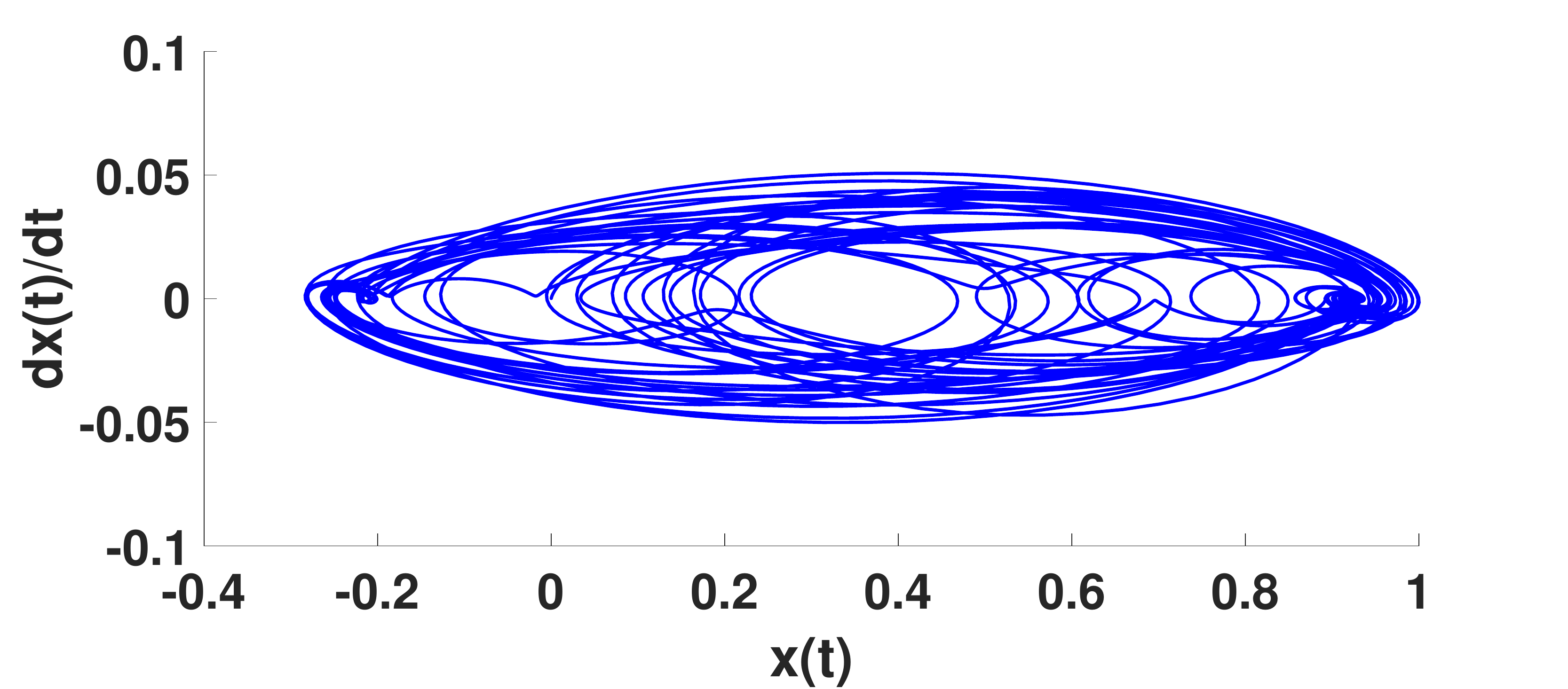}%
}\hfill
\subfloat[]{%
  \includegraphics[width=60mm,scale=0.3]{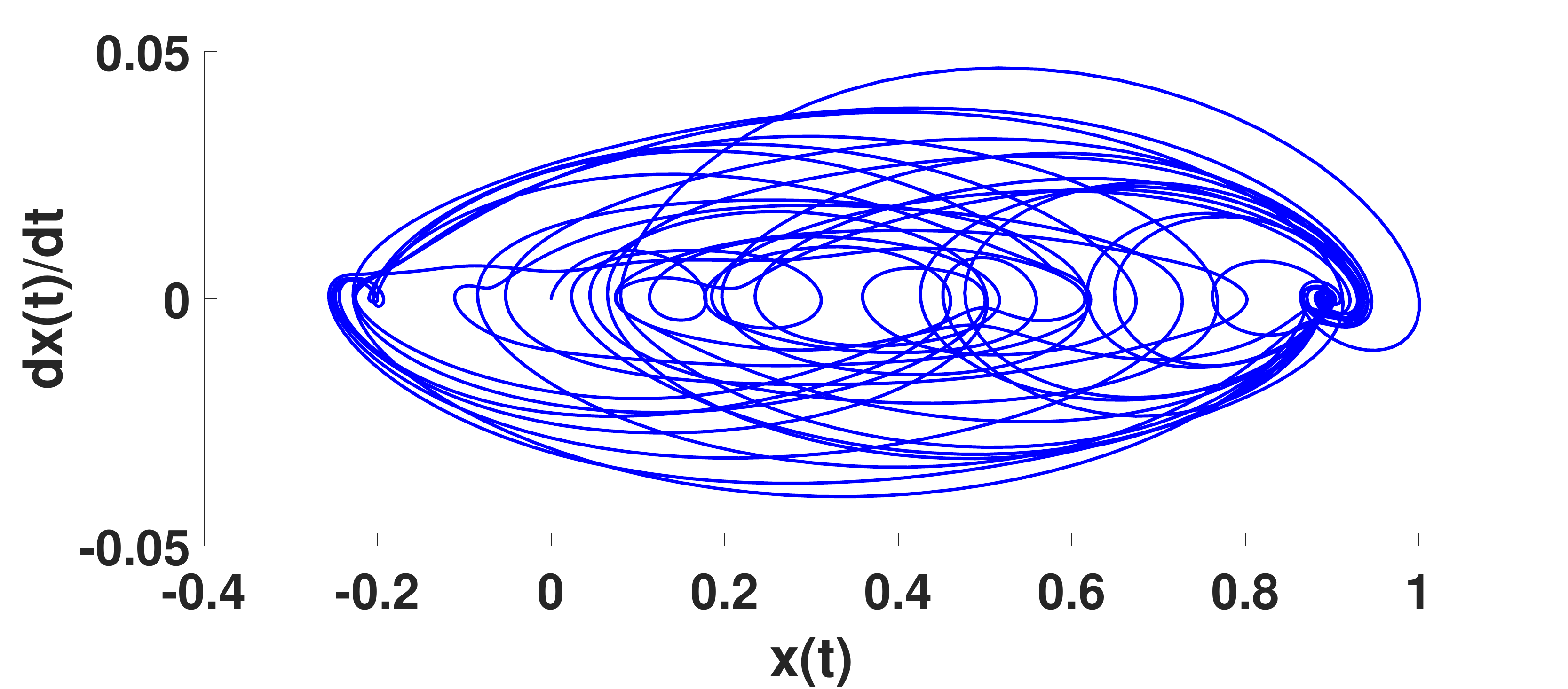}%
}
\subfloat[]{%
  \includegraphics[width=60mm,scale=0.3]{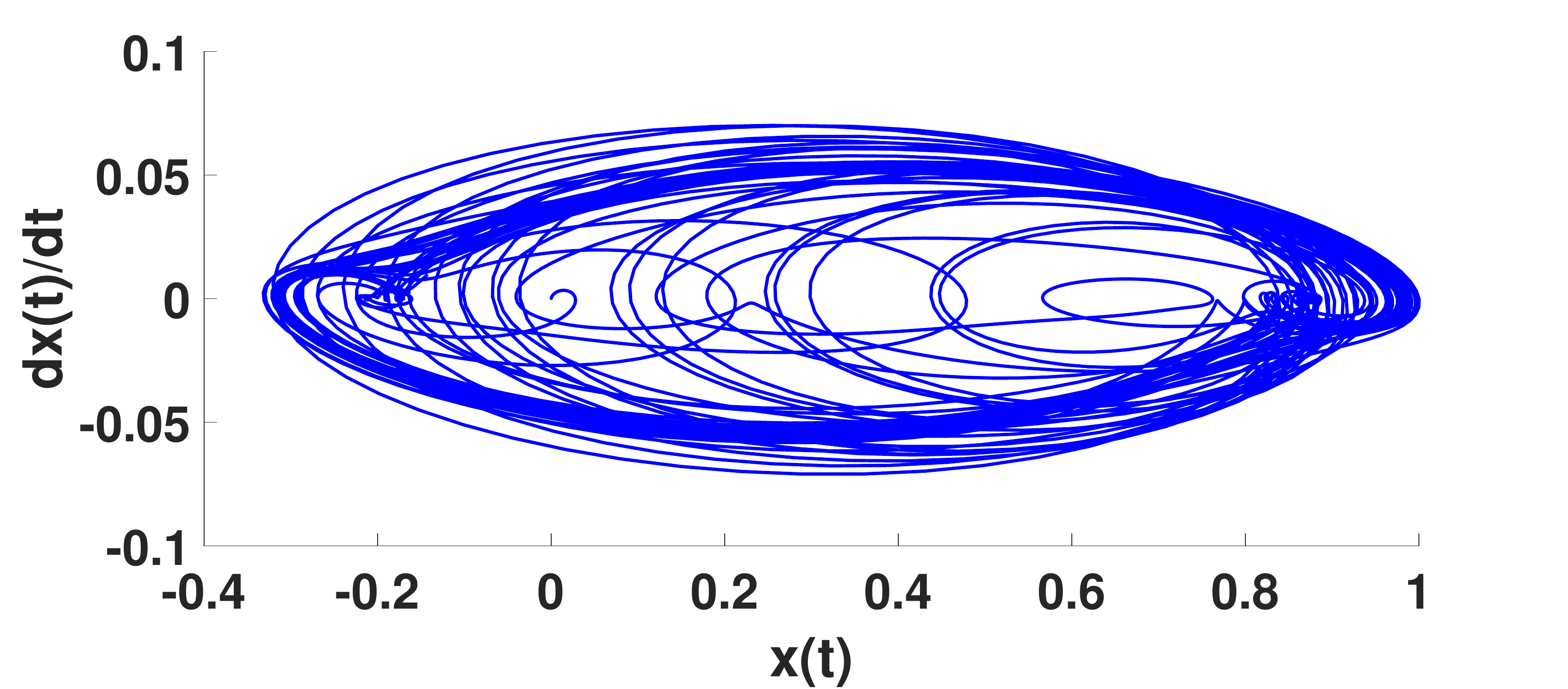}%
}
\subfloat[]{%
  \includegraphics[width=60mm,scale=0.3]{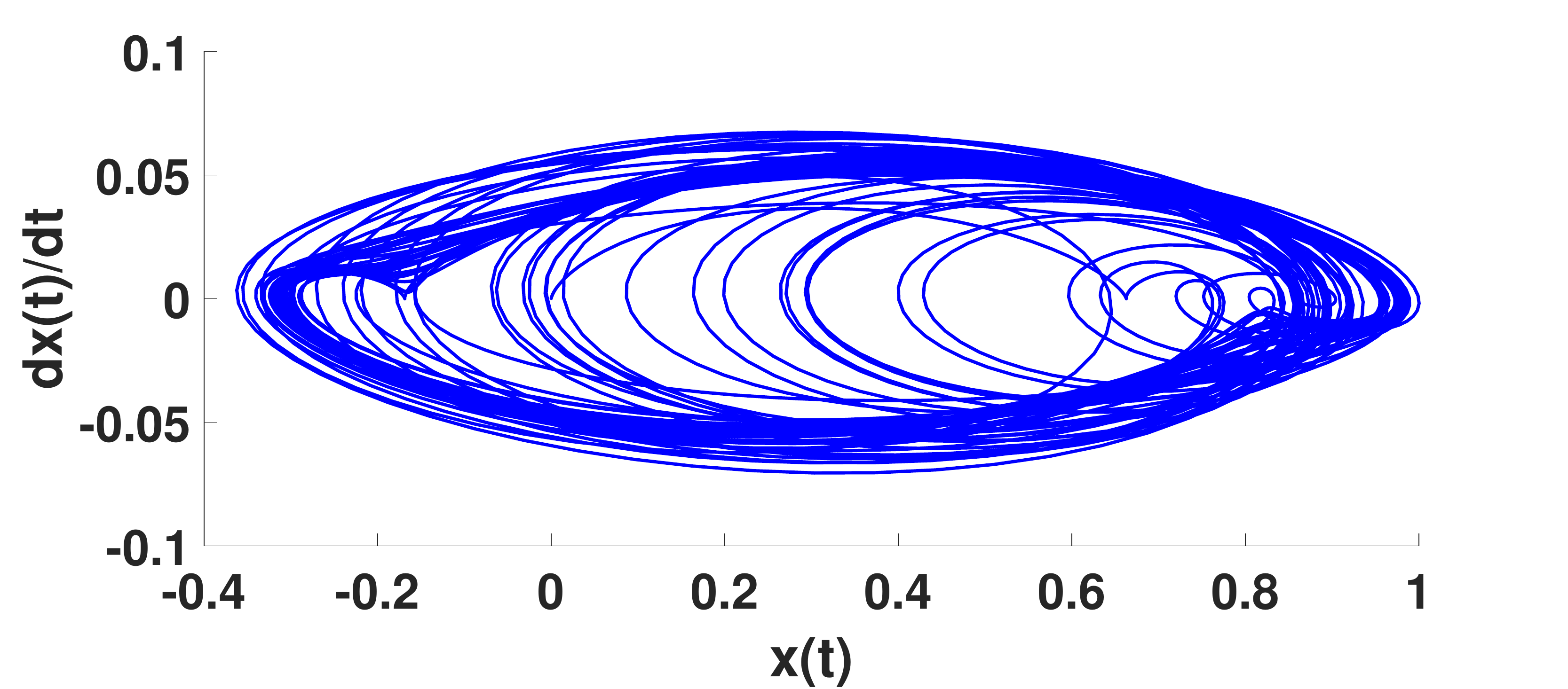}%
}\hfill
\caption{(a),(b) are phase-space images generated from the 50th and 350th rows of Image 1, (c),(d) are phase-space images generated from the 50th and 350th rows of Image 4, (e),(f) are phase-space images generated from the spatial series of Image 12.}
\end{figure*}
 The theory of chaos in a finite-dimensional dynamical system such as in nonlinear ordinary differential equations~\cite{Ref. 5, Ref. 6, Ref. 7, Ref. 8} has been extensively studied. However, exploration of deterministic chaos in partial differential equations (PDE) offer challenging problems~\cite{Ref. 9, Ref. 10, Ref. 11}. An investigation on the simplest nonlinear PDE reveals that the Kuramoto-Sivashinsky is the simplest known PDE with quadratic or cubic nonlinearity and periodic boundary conditions to show chaos~\cite{Ref. 12, Ref. 13}. We have used a model equation known as Swift Hohenberg (SH) equation of the fourth order~\cite{Ref. 14} parabolic type that is famous for its use in pattern formation~\cite{Ref. 15} due to finite length instability, such as in Rayleigh-B\'enard convection~\cite{Ref. 16, Ref. 17}. Due to the vast applicability of the SH equation, it is represented as a generic model for the spatiotemporal dynamics of spatially extended systems~\cite{Ref. 18}. 
It is well known that ordinary complicated mixing show multifractality (for two-components or multi-components fluids)~\cite{Ref. 19}. Being a model pattern forming equation for simple and complex fluids, the idea can also be applied to the patterns generated by SH equation~\cite{Ref. 20} if the evolution dynamics show scale invariance with the existence of self-regulatory internal mechanisms that drive the system spontaneously to a statistical stationary state. The motivation for choosing SH model lies in the statements of Cross' and Hohenberg's work~\cite{Ref. 21} where they have emphasized the omnipresent presence of self-similar systems over a wide range of parameter in natural nonequilibrium systems carefully tested by Bak et al.~\cite{Ref. 22}. Also, spatio-temporal chaos rarely exhibits scaling behavior and that occurs only due to parameter tuning. Considering such  occurrences of multi-fractal scaling properties conditioned to parameter tuning~\cite{Ref. 23, Ref. 24, Ref. 25, Ref. 26}, we have considered SH equation to be a potential candidate for exhibiting multi-fractality in the ensuing steps of the investigation. 

Here, the patterns are nothing but the solution of the SH equation which has been represented using MATLAB colorbar. Time evolution is generated by obtaining snapshots at different time instants whereas space evolution is determined by  varying the parameter values. The color difference between each unit pattern points to a different intensity value which in turn serves as the spatial series data. The data exhibits a robust power law behavior where the log-log plot shows two scaling regions with two distinct slopes. The range of power law scaling factor, $\beta$ varies from 1.3 to 2.8. 

In Section II, we describe the SH equation and the method used to numerically solve it. Section III explains the method used to extract the data points for the spatial series generated from the patterns formed by the SH equation. Section IV shows the spatial series and its corresponding phase space plots. In Section V, we present results of the long range correlations of the chaotic data obtained by Detrended fluctuation analysis (DFA) and Rescaled range analysis (R/S). In section VI, power spectrum by fast Fourier transform and its log power-log frequency plots are shown. The presence of more than one slope in power law diagram confirms multifractality in the SH equation. In Section VII, we have presented the summary of the results~\cite{Ref. 27, Ref. 28}.

\section{\label{sec:level1}Patterns generated from the numerically approximated solutions of the SH Equation}
We have used the reduced third order form of  the SH equation where z variable has been omitted.  
The Swift Hohenberg equation~\cite{Ref. 29} is described as
 \begin{equation}
 \Psi = \epsilon\psi - (\nabla^2+1)^2\psi +v\psi^2-\psi^3 \hspace{0.2cm} where \hspace{0.2cm} \Psi = \frac {\partial \psi}{\partial t}
 \end{equation}

In this equation, $\epsilon$ is the real bifurcation parameter whereas $v$ is the strength of the quadratic equation. Together with $\psi^3$  the term $v\psi^2-\psi^3$ makes the nonlinear term in the equation. Here, the $\nabla^2$ term in the equation is called a spatial derivative complex. Using the Fourier spectral method~\cite{Ref. 30} and Exponential Time Differencing method~\cite{Ref. 31} of order 2 the numerically approximated solutions of the SH equation have been obtained. The Fourier spectral method has been used to transform the PDEs into ODEs by calculating the Fourier modes. After that, the ODEs are solved using the Exponential Time Differencing method (ETD2). On applying these methods to our equation, a matrix containing the intensity for the x-y plane is generated. The entire simulation has been carried out using MATLAB~\cite{Ref. 32}. At t=0, the initial condition is chosen to be a random matrix.   
 
 \begin{equation}
 \Psi(\textbf{X}(t)) = \Psi(\textbf{X}(0))
 \end{equation}
 
$\textbf{X}$ signifies a two dimensional space i.e. x and y space plane. Here $\Psi(\textbf{X}(0))$ is a random number from the interval $-\sqrt{\epsilon}  \leq \Psi(\textbf{X}(0)) \leq \sqrt{\epsilon}$. Here, $v$ and $\epsilon$ values are the only control parameters which can be varied to observe spatial changes. To notice the time evolution, we have recorded the image changes at t=5 seconds, t=10 seconds, t=15 seconds, and t=20 seconds. For our simulation we have chosen 256 x 256 grid points with a time step of 0.1.

Images 1-16 of Figure 1 are the spatio-temporal evolution of SH equation with a defined initial condition and a set of parameters i.e. $v$ and $\epsilon$ and t.  The red disks represent the highest image intensity value (portrayed by jet colorbar, MATLAB) whereas blue represents the lowest value. The next section has been dedicated to explaining the conversion of image intensity to spatial data.

\begin{figure*}[h]
\centering
\includegraphics[width=170mm,scale=0.3]{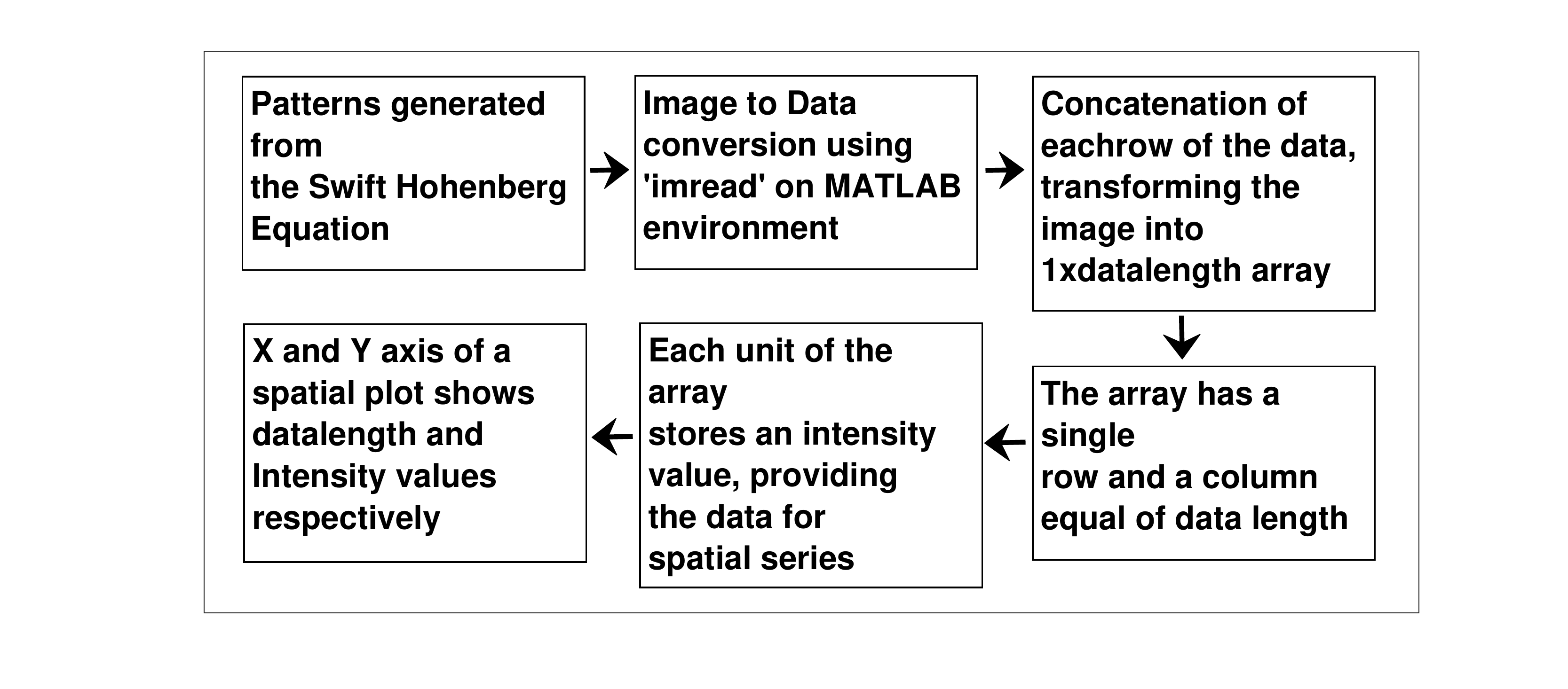}%
\caption{A flowchart description of the image to data conversion method}
\end{figure*}
\begin{figure*}
\subfloat[]{%
  \includegraphics[width=60mm,scale=0.3]{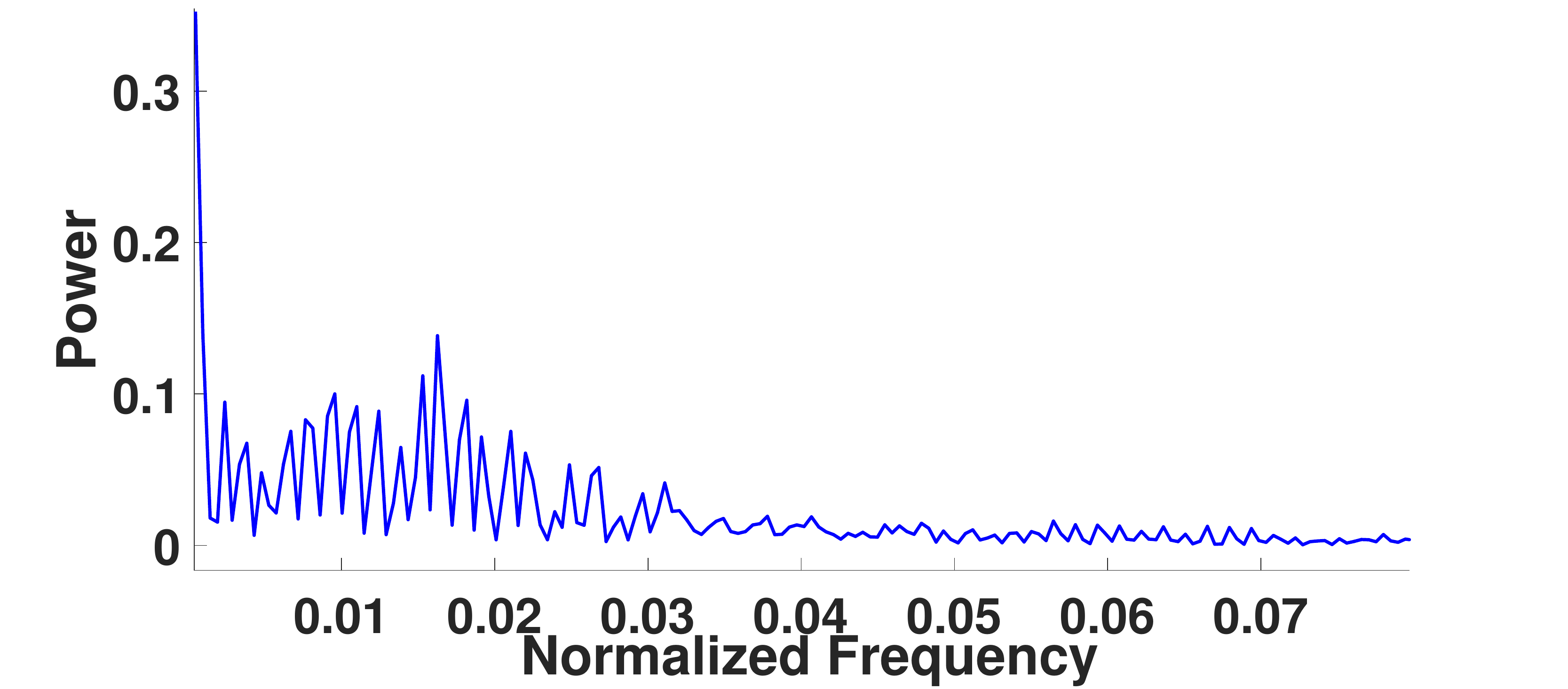}%
}
\subfloat[]{%
  \includegraphics[width=60mm,scale=0.3]{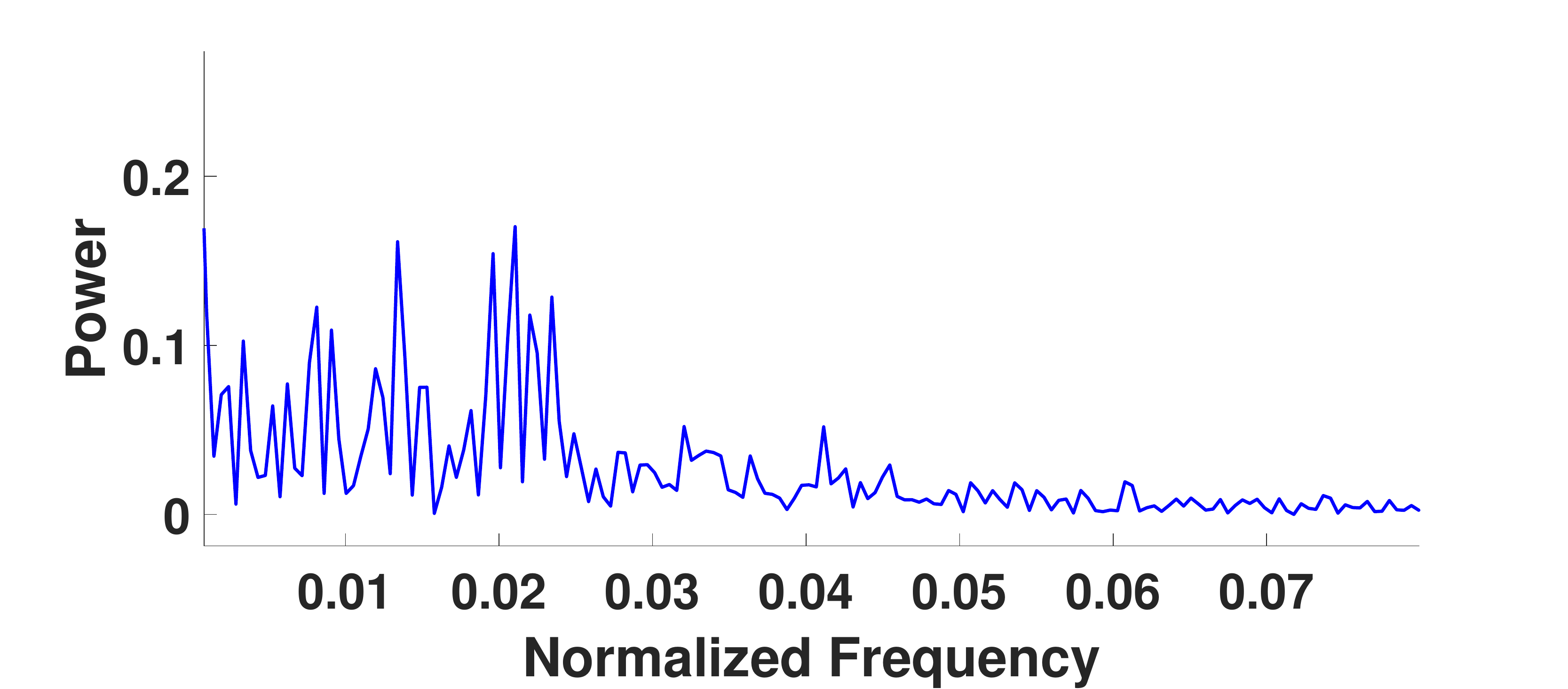}%
}
\subfloat[]{%
  \includegraphics[width=60mm,scale=0.4]{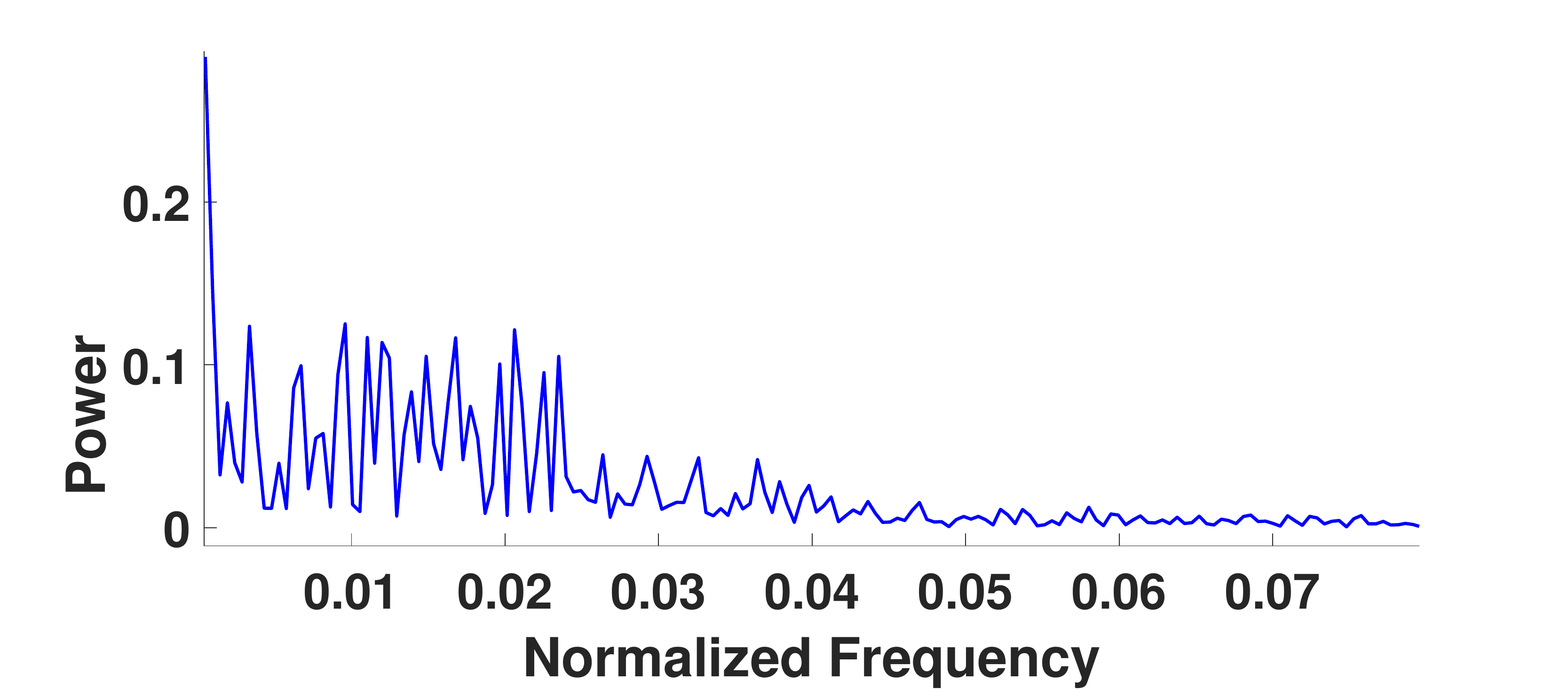}%
}\hfill
\subfloat[]{%
  \includegraphics[width=60mm,scale=0.3]{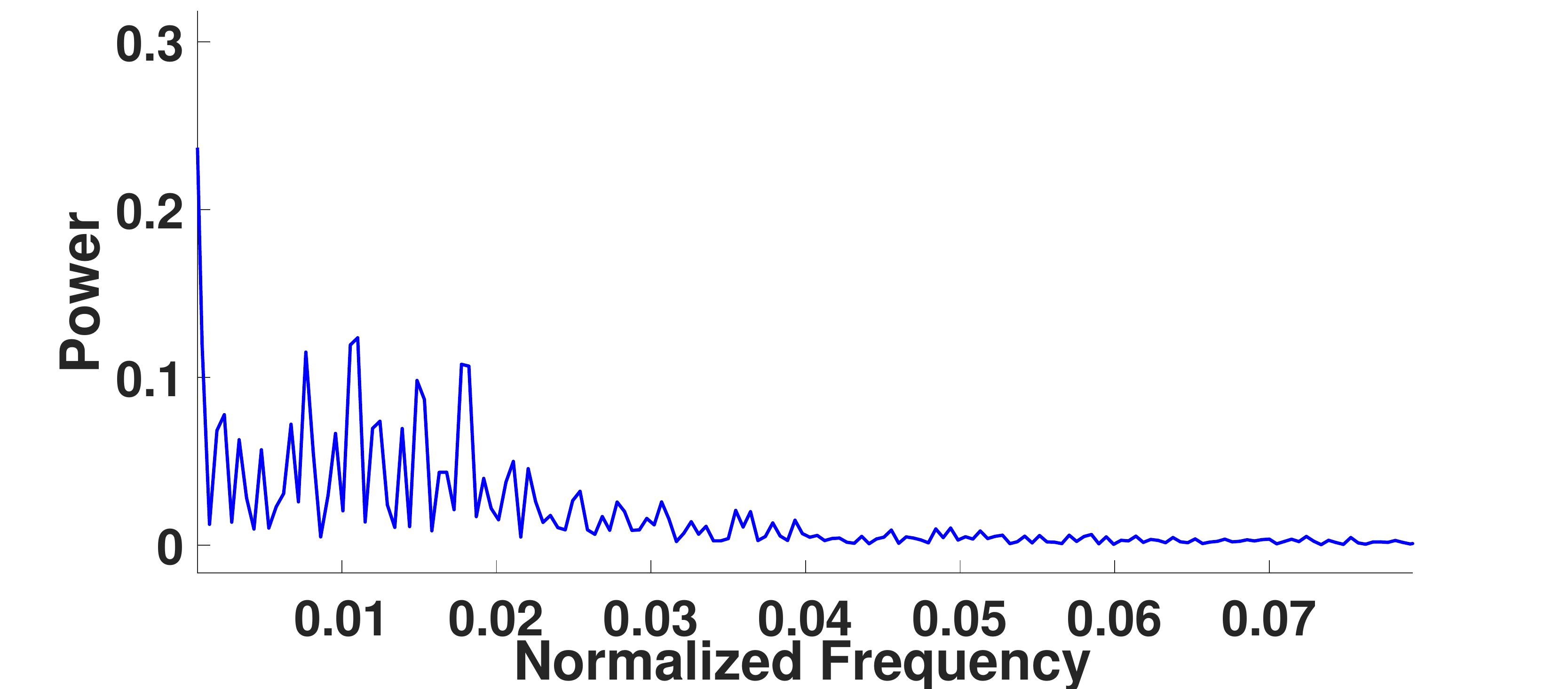}%
  }
\subfloat[]{%
  \includegraphics[width=60mm,scale=0.3]{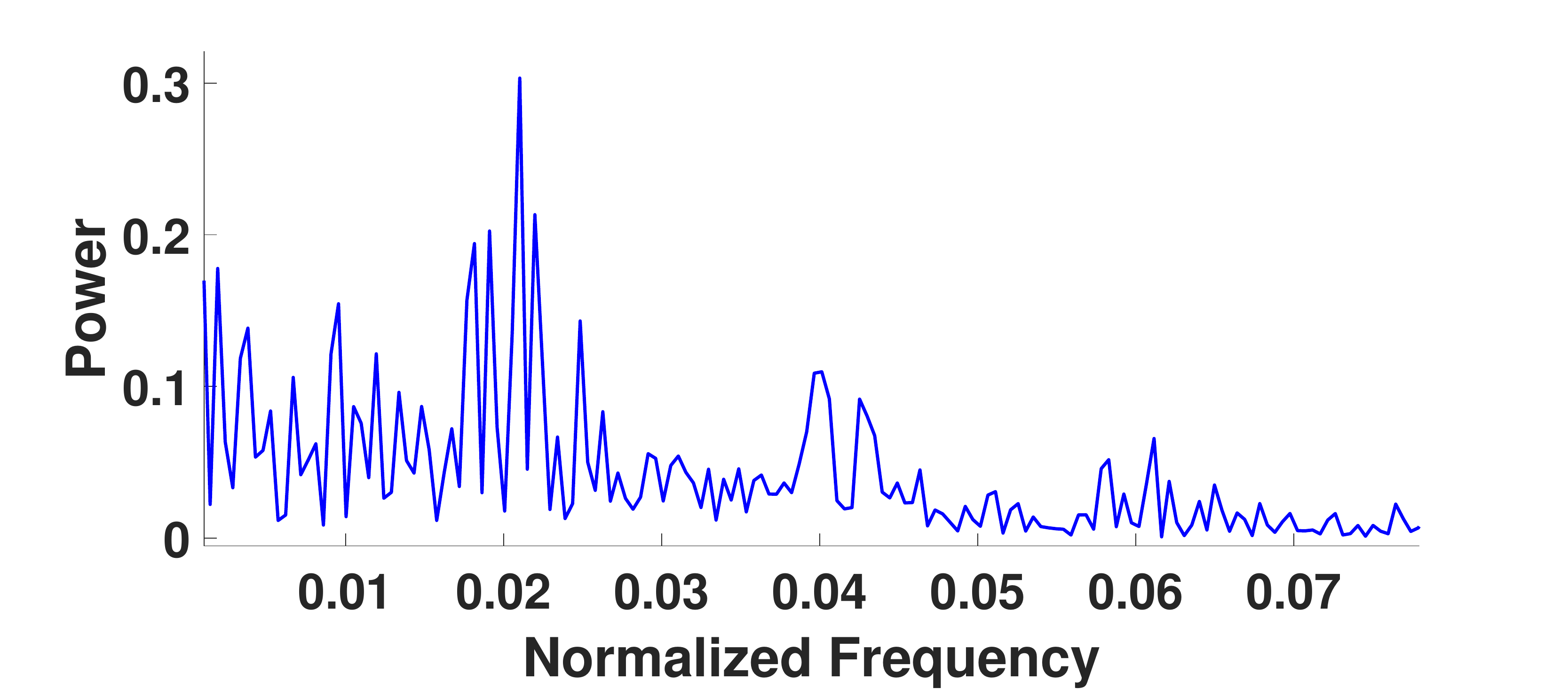}%
}
\subfloat[]{%
  \includegraphics[width=60mm,scale=0.3]{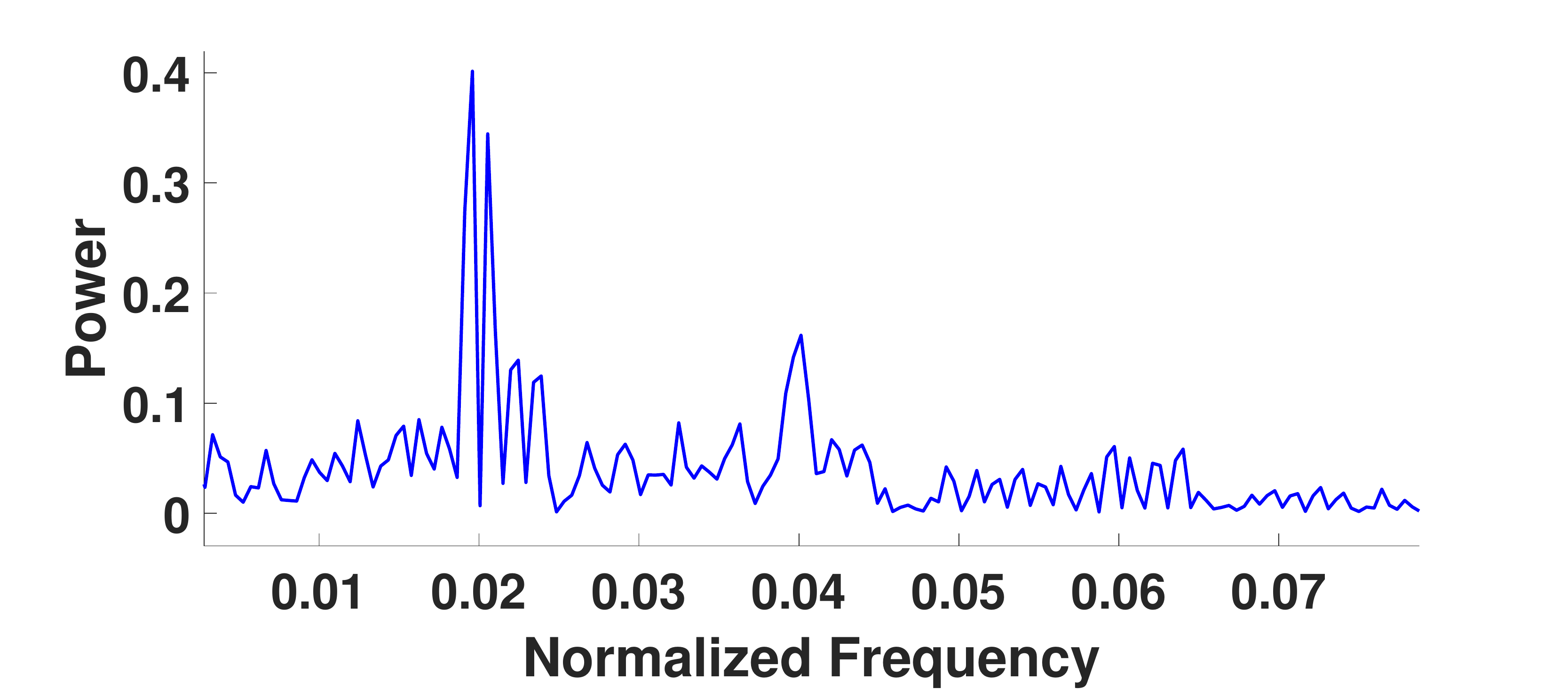}%
  }\hfill
\caption{(a),(b) are power spectrum images generated from the 50th and 350th rows of Image 1,(c),(d) are power spectrum images generated from the 50th and 350th rows of Image 4, (e),(f) are power spectrum images generated from the spatial series of Image 12.}
\end{figure*}




\section{\label{sec:level1} Explanation of image intensity to spatial series conversion}
Initially, we have obtained numerically approximated solution from the SH equation and have reconstructed the solutions to images. This makes the spatio-temporal evolution and its pattern forming characteristics to become visually perceptive. One of the basic characteristics of image processing in MATLAB is that an image can be converted into a matrix and the opposite also holds true. This unique convertibility has been used here to extract back the spatial data from the images where each pixel (color grain) corresponds to a lower or a higher value of the matrix. Here, the solution of the equation is the intensity values of images represented by $\psi$. For this simulation, “Jet” colorbar has been chosen to represent the patterns in Matlab figures. The key significance of colorbar is its usability in defining color scheme for many types of visualizations, such as surfaces and patches. The intensity of the images is represented in such a way that a redder appearance shows a higher intensity value whereas a bluer appearance shows a lower intensity value. Accompanying colorbars with Figure. 1 describe the entire range.\\

Seemingly redundant, this double conversion (data to image and image to data) has its use in experimental systems, where the spatial data is not known. The technique can be very convenient for obtaining information directly from experimentally generated 3-dimensional images such as pattern formation in developmental biology where intensity of a picture is of great importance along with 2 space dimensions. Once converted from patterns,the data has multitude of applications including their use for traditional nonlinear time series analysis. We are proposing this method for medical imaging, vegetation patterns, Turing patterns, reaction-diffusion systems and any other pattern formation variants where space, time or space-time based evolution is experienced. A similar but experimental approach i.e. image to data conversion for studying chaotic dynamics of microscale patterns formed due to demixing of two fluids has been reported in~\cite{Ref. 33} where they have noticed a transition from long range correlation to long range anti-correlation as the system progresses towards the final state.

On the next step we have applied the nonlinear time series analysis on the extracted spatial series data and have estimated the fractal dimension of the system. The two dimensions of the patterns i.e. the abscissa X and ordinate Y of the patterns are the two space variables x and y. Each unit intensity of an image is the solution of SH equation for two space variables namely x and y at a certain time where the two varying parameter namely $v$ and $\epsilon$ is assigned to a specific value. The method proposed in this work is a simple image to matrix conversion method, where these color intensity values represent specific colors as given in the colorbar. Now, each row of the image is taken and then concatenated with the next immediate row. Once the entire 2D image is converted into a single row data, it is plotted as a spatial series where y-axis represents the intensity values and x axis represents the data length. A flow chart (Figure 4), explaining the method is also attached for convenience.

\section{\label{sec:level1} Phase space analysis and Lyapunov exponents calculation of spatial series}

Since we have obtained a set of pattern forming images depending on varying time and parameter values, we have taken a methodical approach to choose three images out of that set to be used further for our analysis. The process of selection is made in such a way, that the change of pattern is noticed due to the change of $\epsilon$ and time. For this experiment, we have taken Image 1 as our initial sample which has $v = 0.1, \epsilon = 0.3, t = 5$ seconds. Along with it we have taken Image 4 ($v = 0.1, \epsilon = 0.3, t = 20$ seconds - time increased), and Image 12 ($v = 0.1, \epsilon = -0.3, t = 20$ seconds - $\epsilon$ and time increased) as a representation of spatio-temporal evolution due to the difference in time and one of the varying control parameter $\epsilon$. After acquiring the entire spatial series, we have looked at the 50th row data and 350th row data of each of the aforementioned images. 


\subsection{\label{sec:level2}Spatial series formation of SH solution from computationally generated images }

 The intensity profile of the solution depicted in images are converted to data points that generated a specific row vs column matrix. In this paper, we have used a direct approach to convert the image-generated matrix to a spatial series by simply translating data points into a 2D spatial plot. If the matrix has m columns and n rows then it is first converted into a single row containing $n\times m $ columns i.e. 1D array by simply removing the rows from second row onward, until the $n^{th}$ row and concatenating them with the first row. When plotted (Figure. 2), the result is a spatial series with image intensity solution in its Y axis and data length in its X axis. From the plots it is apparent that the data do not hold a periodic nature and gives a typical random appearance. However, apparent randomness is not enough to justify the data and therefore the data has been used for further analysis.

\subsection{\label{sec:level2}Phase space analysis }

In dynamical systems, phase space projection plays a decisive role in determining all possible states of a system. We have obtained a normalized differential phase space plot via plotting a graph between spatial series data points versus its derivative. 

The phase space projections represented in the Figure 3 show trajectories of a dynamical system for different row data. The data used is filtrated from noise by making a use of low pass Butterworth filter. The noise is introduced when the spatial series data is differentiated. This happens because of the sharp changes in the spatial series data where the derivative values can not be computed. The low pass Butterworth filter removes those high frequency fluctuation which are redundant for our calculation method.\\

One of the primary points to be noted from the phase space results is the position of the periodical trajectories or orbits. The trajectory encircles around one basin of attraction~\cite{Ref. 34} and suddenly it starts to circle another.
The trajectories represent a very dense pathway which does not follow the same route again however co-exist with its neighbor in a close distance making the area thicker in appearance. This corroborates with the concept of an attractor which by definition means the set of states, invariant under dynamics, towards which neighboring states in a given basin of attraction asymptotically approach in the course of dynamic evolution.~\cite{Ref. 35} As commonly known, this behavior is natural to chaotic data sets. Although it does not offer any specific information about the dataset but significantly affirms the nature of the data being chaotic.
One of many other popular approaches to detect chaos has been assessing whether the largest Lyapunov exponent ($\lambda$) gives a positive value. The Lyapunov exponent of a dynamical system is a quantity that characterizes the rate of separation of infinitesimally close trajectories. If $\lambda > 0$, the nearby trajectories separate exponentially, determining chaos. Using the TISEAN package\cite{Ref. 36} based on Rosenstein algorithm ~\cite{Ref. 37} we have derived the largest Lyapunov exponent values for the specific set of images, for Image 1, Image 4 and Image 12. The two key factors that can manipulate the largest Lyapunov exponent values are embedding dimension and time delays. Even though determining the correct embedding dimension is an open challenge, we have used False Nearest Neighbor ~\cite{Ref. 38} method to find the embedding dimension values. We have used Mutual Information ~\cite{Ref. 39} technique to calculate the time delay. Out of the entire set represented in TABLE II the largest LLE comes to be 8.4282.

\section{\label{sec:level1}Comparing and quantifying long range Correlation using Detrended fluctuation analysis and Rescaled range analysis}

Below, both the standard methods such as Detrended fluctuation analysis and Rescaled range analysis, known for calculating Hurst exponent values, have been used. In this paper, the range of Hurst exponent values obtained have played as primary determinant to signify the presence of long range correlation in the data. 


\subsection{\label{sec:level2} Detrended fluctuation analysis (DFA)}
Detrended fluctuation analysis~\cite{Ref. 40} tool, applied over the same spatial data, is a method to investigate the statistical self-similarity of a signal and has been proven to be useful in revealing the extent of long-range correlations in time series~\cite{Ref. 41, Ref. 42}. A general feature experienced in all self-similar systems is that the temporal and/or spatial power-law correlations extend over several decades where intuitively one may anticipate that the physical laws would change drastically. 

A simplified and general definition characterizes a time series as stationary (in this case spatial series) when the mean, standard deviation and higher moments, as well as the correlation functions are invariant under time translation. Signals that do not obey these conditions are non stationary. However the upper mentioned Detrended fluctuation analysis allows the discovery of intrinsic self-similarity hidden inside an apparently non-stationary time series or spatial series. The method is a technique for measuring the same power law scaling observed through R/S analysis, explained later but it has been used specifically to address non stationaries. DFA is commonly known as enabling correct estimation of the power law scaling (Hurst exponent) of a systems' signal in the presence of (extrinsic) non stationaries while eliminating spurious detection of long-range dependence.

The algorithm can be explained in four steps as follows:- \\
1. \textbf{Integration:} The spatial series data $x(i)$ is shifted by the mean $\bar x$ and cumulatively summed.

\begin{equation}
 y(i) = \sum_{j=1}^{i} (x(i)-\bar x)
\end{equation}

2. \textbf{Segmentation and Fitting:} The obtained sum $y(i)$ is then segmented into boxes of $n$ equal length. Then a local trend is calculated by fitting polynomial $y_{n}(i)$ to each box. 

3. \textbf{Detrending:} The detrended series is then calculated by the following equation.

\begin{equation}
 Y_n = (y(i) - y_n(i))
\end{equation}

4. \textbf{Deriving DFA fluctuation function:} The root mean square deviation from the trend, i.e. the fluctuation function is calculated following the equation below. 

\begin{equation}
 F(n) = \sqrt{\frac{1}{n}\sum_{i=1}^{N} (Y_n (i))^2}
\end{equation}

Where $N$ = entire length of the series

Following the power law correlation, self affinity of the spatial series can be represented as 

\begin{equation}
 F(n) \propto n^\alpha
\end{equation}

where $\alpha$, the scaling constant is obtained via  the slope of a straight line fit to the log-log graph of $n$ vs $F(n)$. For our case, $\alpha = H+1$~\cite{Ref. 43, Ref. 44}.



\subsection{\label{sec:level2} Rescaled range analysis (R/S) }

The rescaled range analysis~\cite{Ref. 45} is a statistical tool to measure the variability of a time series (or spatial series) introduced by the British hydrologist Harold Edwin Hurst (1880-1978)~\cite{Ref. 46}. The R/S analysis determines how the apparent variability of a series changes with the length of the time-period being considered.

Introduced by Hurst, the Rescaled range (R/S) analysis~\cite{Ref. 47} splits a time series (or spatial series) into adjacent windows and inspects the range R of the integrated fluctuations, rescaling by the standard deviation S as a function of window size~\cite{Ref. 48}. Compared to the DFA analysis examined above, the R/S analysis gives attention to the range of signal instead of finding fluctuations around trend. As a result, DFA is suitable to be applicable for nonstationary time series analysis in contrast to R/S.
Here we applied both of the methods to the spatial series data we have received from the n $\times$ m matrix to draw both the DFA and R/S plots.

The R/S analysis can be performed by following the mentioned algorithm below. 
\\

1.  \textbf{Shifting:} The spatial series data is shifted by the mean for $X = X_1,X_2,X_3...X_n$
{\begin{equation}
 Y_t = X_t - \bar X
\end{equation}}
where $\bar X = \frac{1}{n}\sum_{i=1}^{n}X_t$ and $t = 1,2,...,n$
\\

2. \textbf{Series formation:} A cumulative deviate series $Z$ has been calculated 

{\begin{equation}
 Z_t = \sum_{i=1}^{t}Y_i
\end{equation}}
 
3. \textbf{Calculation of Range:}\\
$R(n) = max(Z_1,Z_2,Z_3,....Z_n) - min(Z_1,Z_2,Z_3,...,Z_n)$
\\

4. \textbf{Standard Deviation calculation:} 
The standard devitation can be calculated as follows - 
{\begin{equation}
 S_t = \sqrt{\frac{1}{t}\sum_{i=1}^{t} (X_i - m)^2}
\end{equation}}
where $m = $ mean of $(X_1,X_2,... X_t)$

5. \textbf{Rescaled Range series estimation:} 
The rescaled range analysis can be done as follows - 
{\begin{equation}
 (R/S)_t = R_t/S_t 
\end{equation}}
where $t = 1,2,... n$

The calculated ratio of R/S follows the relation $R/S = d^H$ where $d$ is a constant and $H$ is the Hurst Exponent. The slope of $log(R/S)$ vs $log(n)$ gives the value of $H$. 

From the plot of $logF(n)$ vs $log(n)$ and $log(R/S)$ and $log(n)$, presented in Figure 6, we have performed a straight line  fitting from which the estimated Hurst Exponent ($\alpha$) turns out to be greater than 0.5. It has been noticed that a larger Hurst exponent value shows a stronger trend. According to our result, the $H$ value obtained from DFA stays between $0.5458$ to $0.8108$ whereas the same obtained from RS Analysis stays between $0.5109$ to $0.8118$. Both the results corroborate with the result obtained from power spectral density analysis and further strengthens the appearance of long range correlation~\cite{Ref. 49, Ref. 50}. 


\begin{figure*}
\subfloat[]{%
  \includegraphics[width=60mm,scale=0.3]{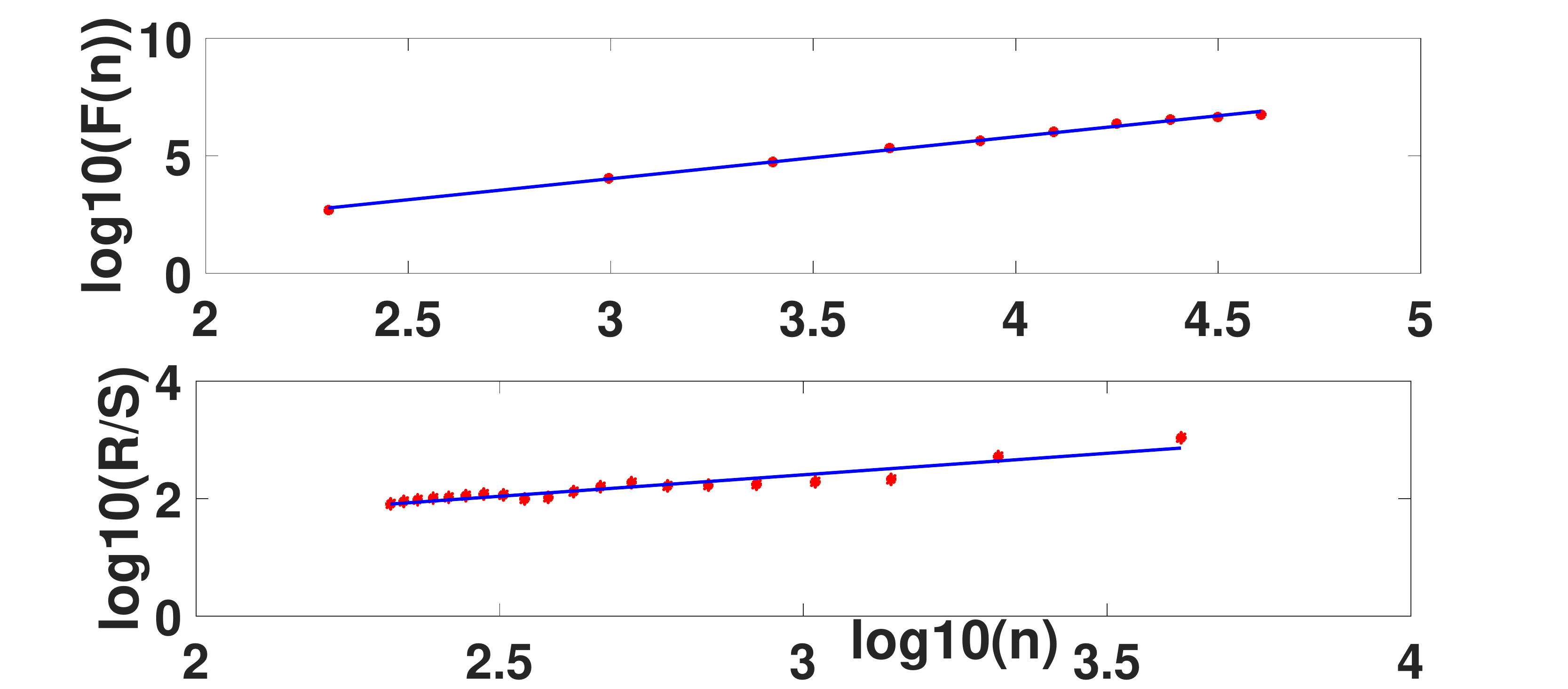}%
}
\subfloat[]{%
  \includegraphics[width=60mm,scale=0.3]{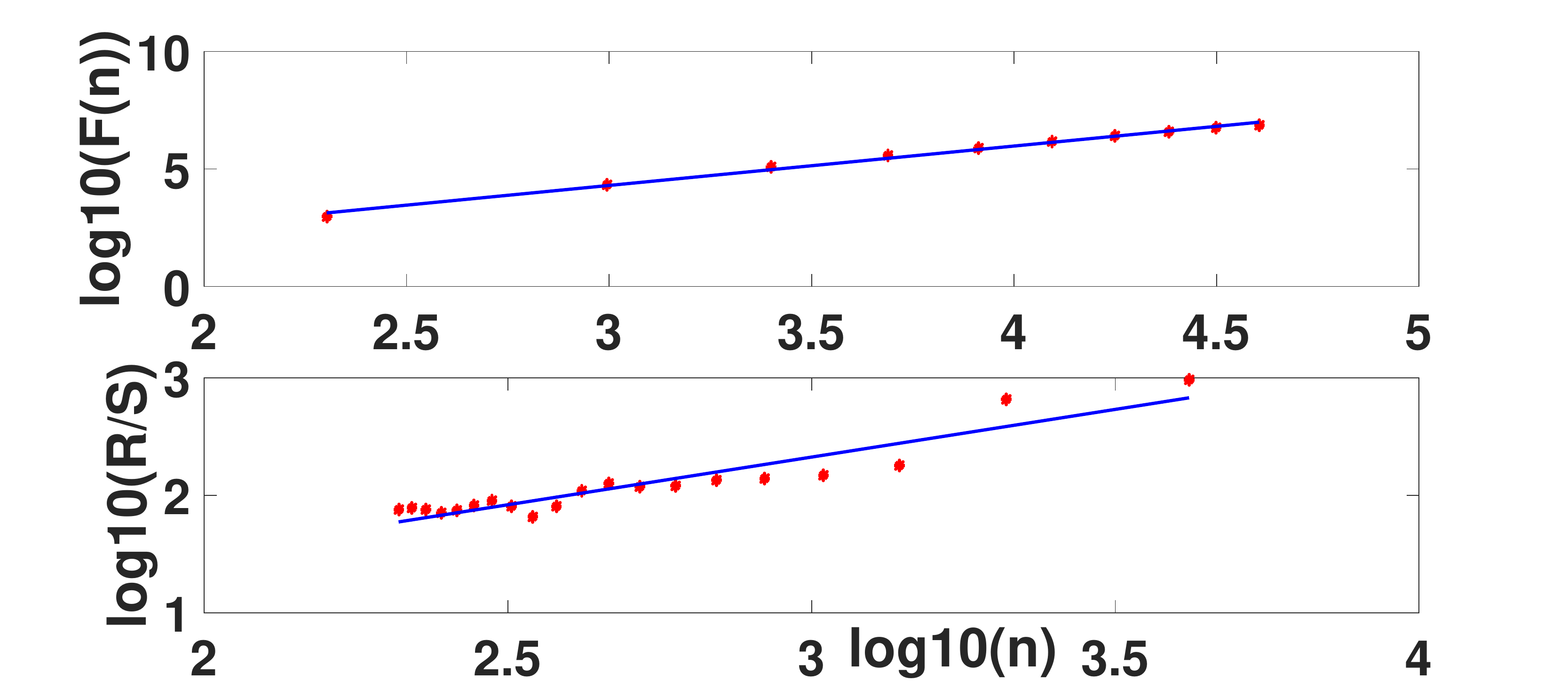}%
}
\subfloat[]{%
  \includegraphics[width=60mm,scale=0.3]{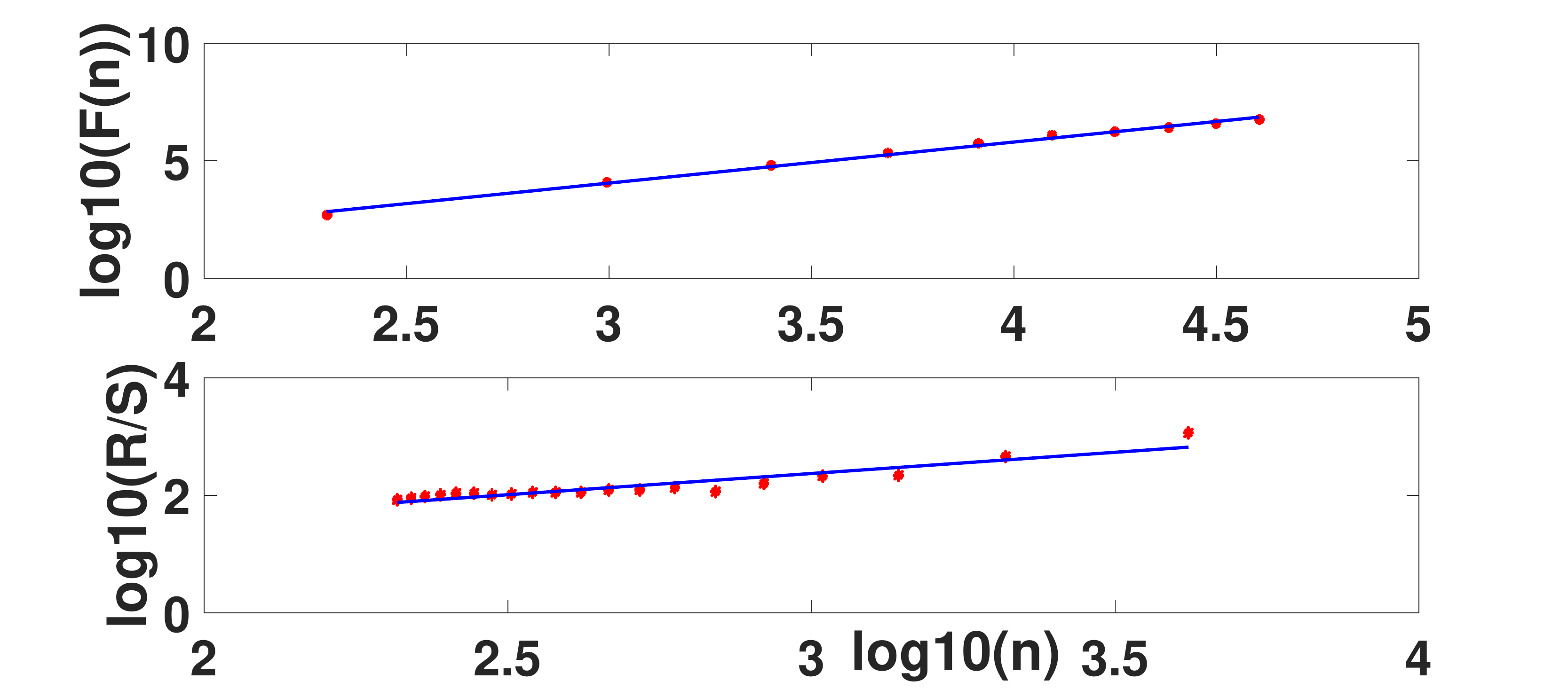}%
}\hfill
\subfloat[]{%
  \includegraphics[width=60mm,scale=0.3]{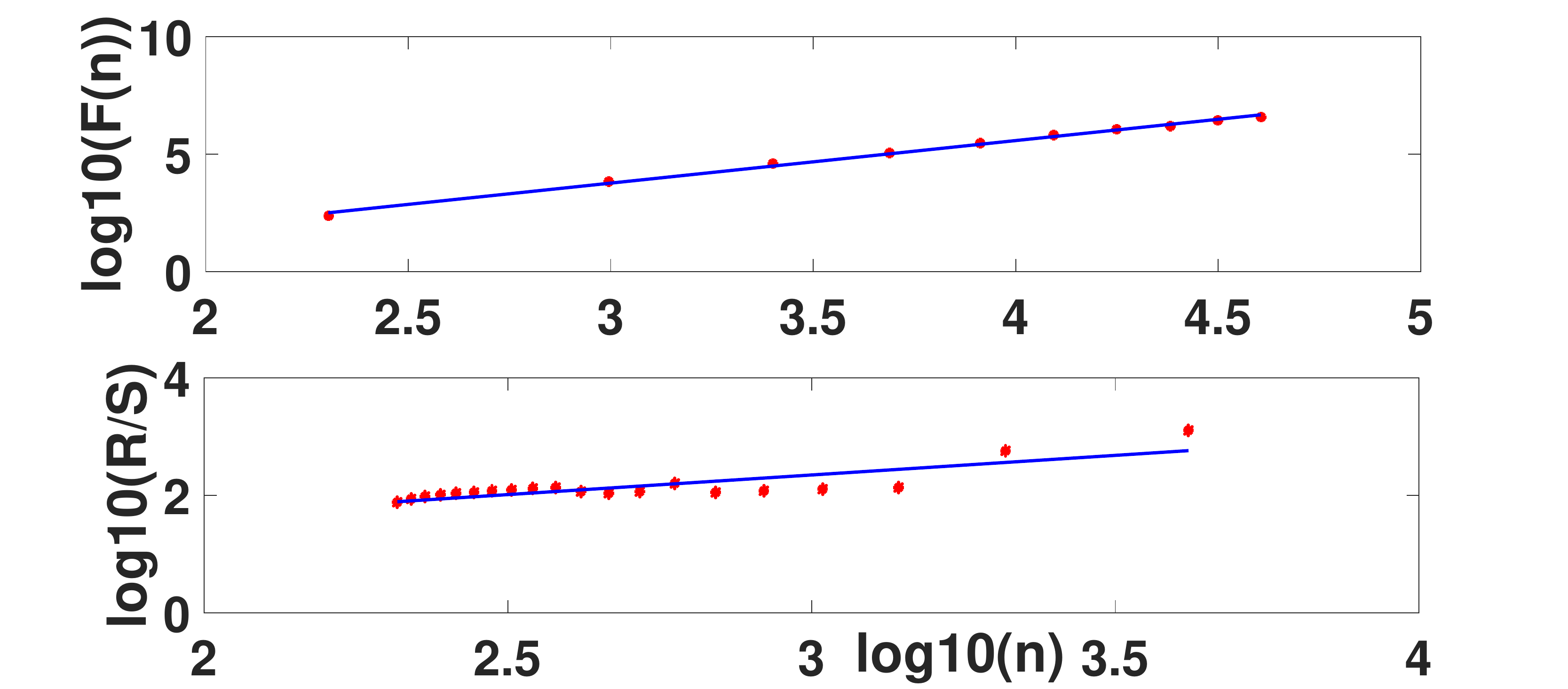}%
  }
\subfloat[]{%
  \includegraphics[width=60mm,scale=0.3]{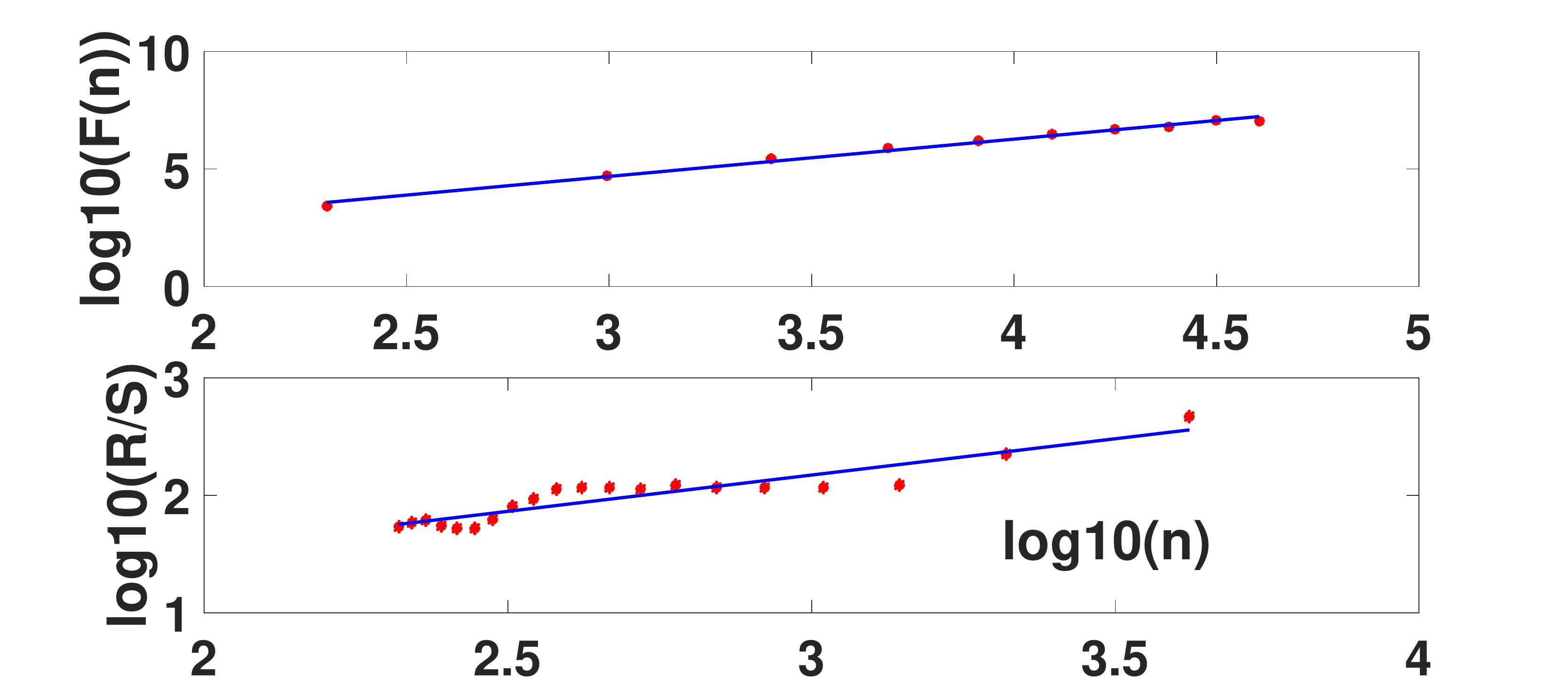}%
}
\subfloat[]{%
  \includegraphics[width=60mm,scale=0.3]{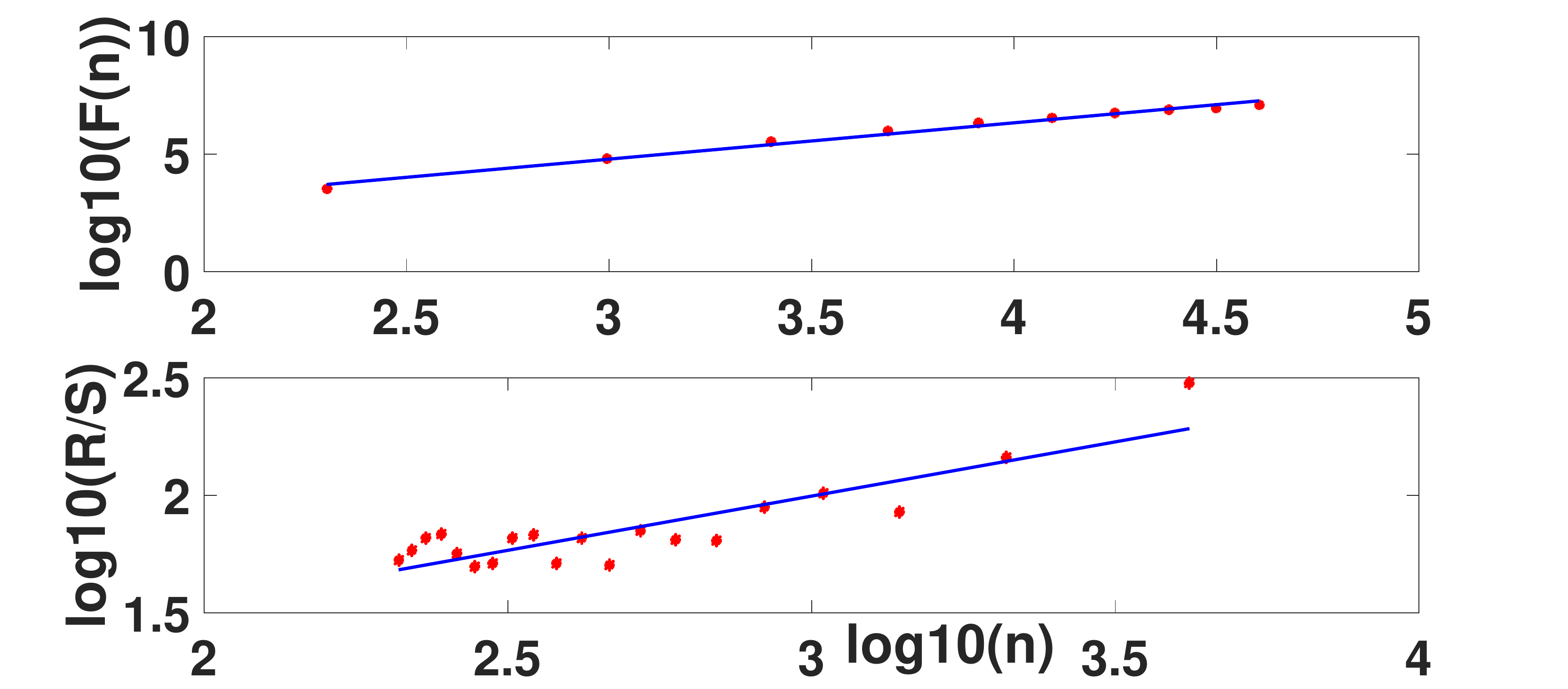}%
  }\hfill
\caption{The log-log plot of DFA and RS analysis are drawn together where blue straight line of DFA and R-S plot are the trend lines or the line of best fit where red stars show the data distribution. (a),(b) pair,(c),(d) pair, (e),(f) pair represent plots for Image 1,Image 4, and Image 12 for 50th row to 150th row  data and 250th row to 350th row.}
\end{figure*}


\begin{figure*}[h]
\subfloat[]{%
  \includegraphics[width=60mm,scale=0.3]{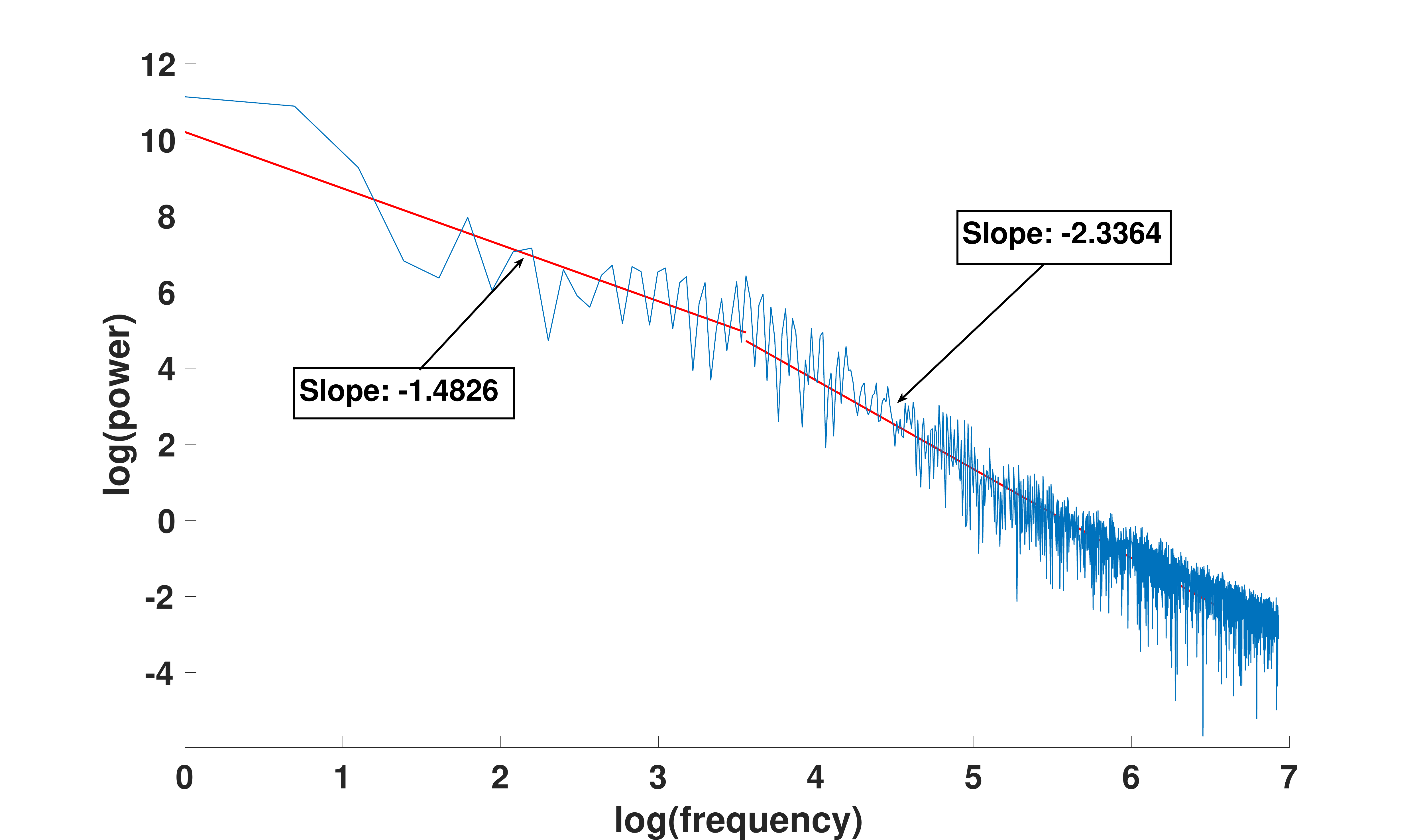}%
}
\subfloat[]{%
  \includegraphics[width=60mm,scale=0.3]{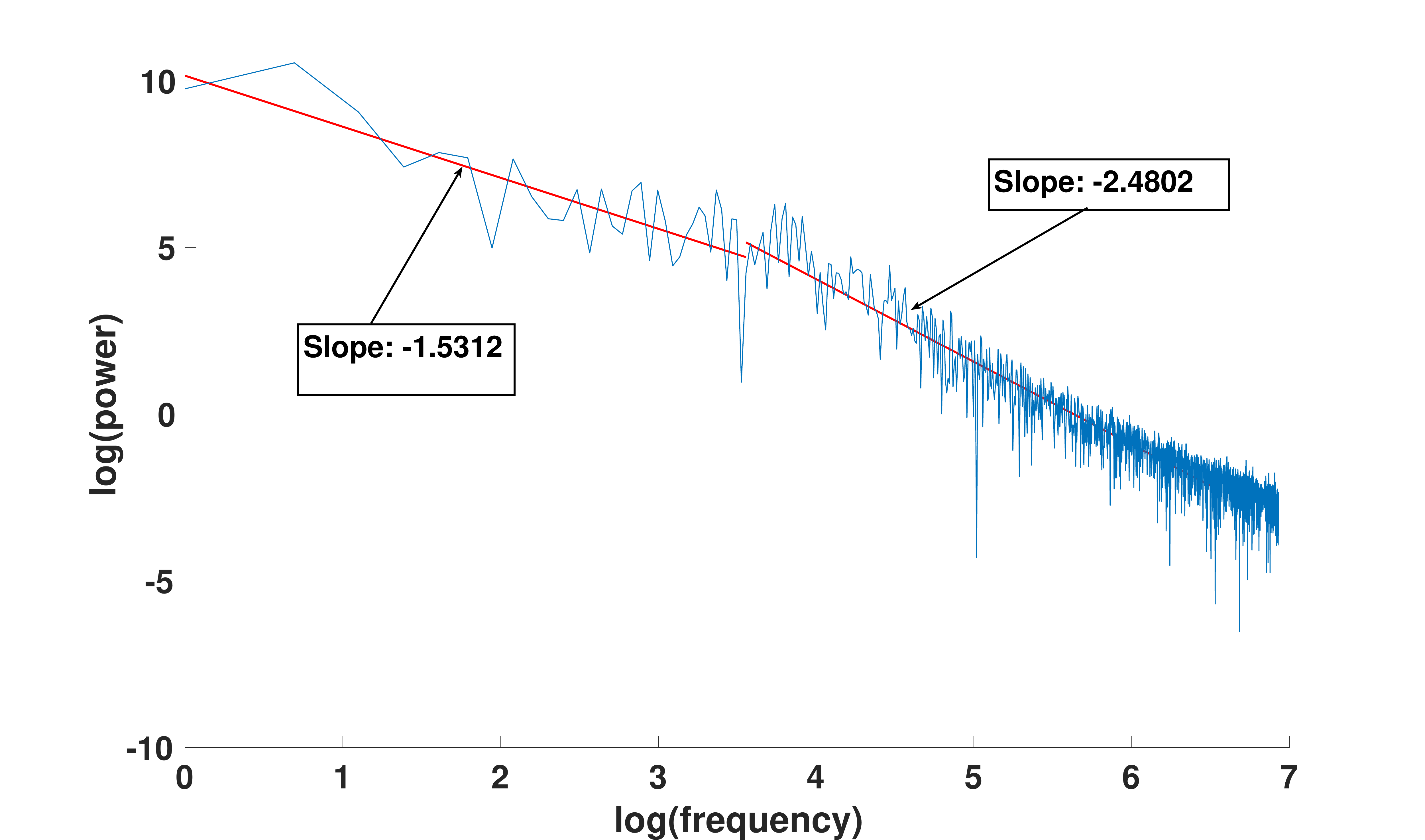}%
}
\subfloat[]{%
  \includegraphics[width=60mm,scale=0.3]{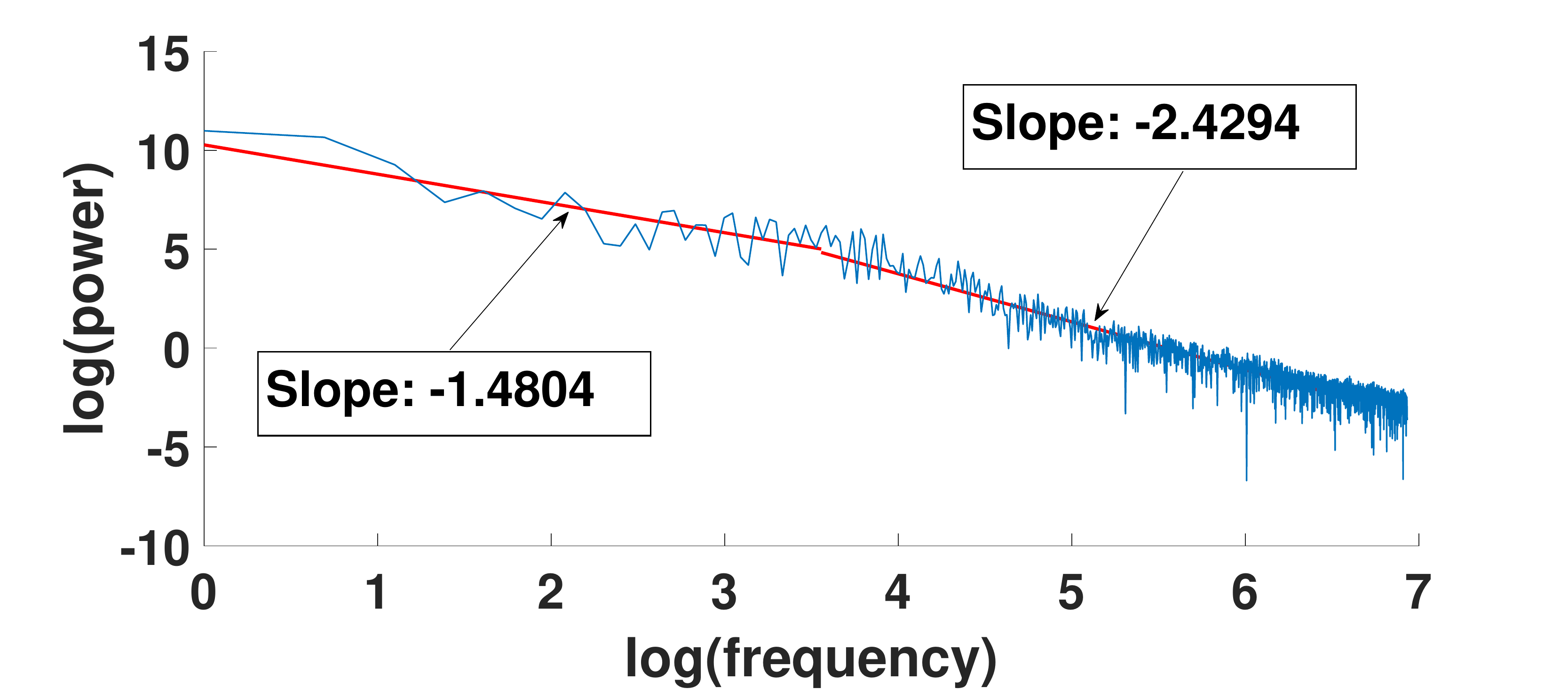}%
}\hfill
\subfloat[]{%
  \includegraphics[width=60mm,scale=0.3]{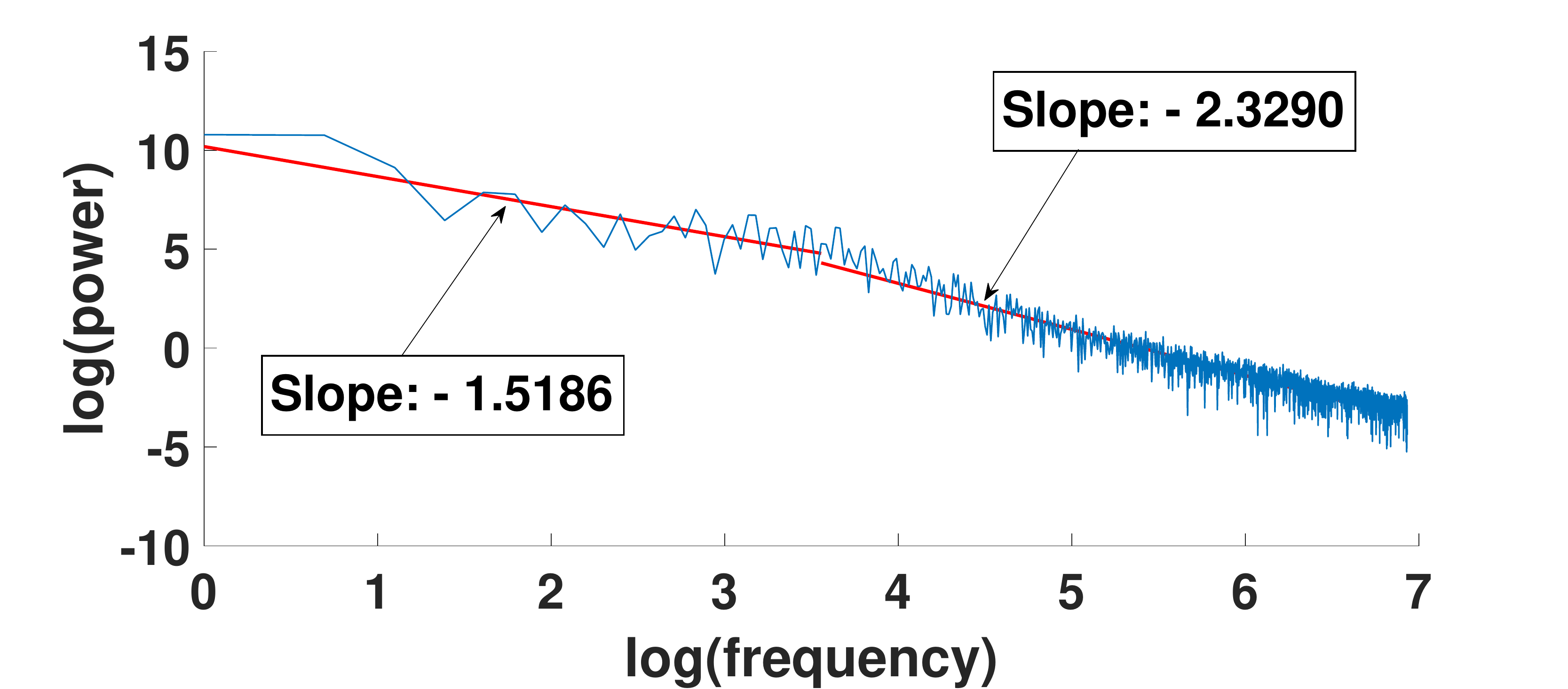}%
  }
\subfloat[]{%
  \includegraphics[width=60mm,scale=0.3]{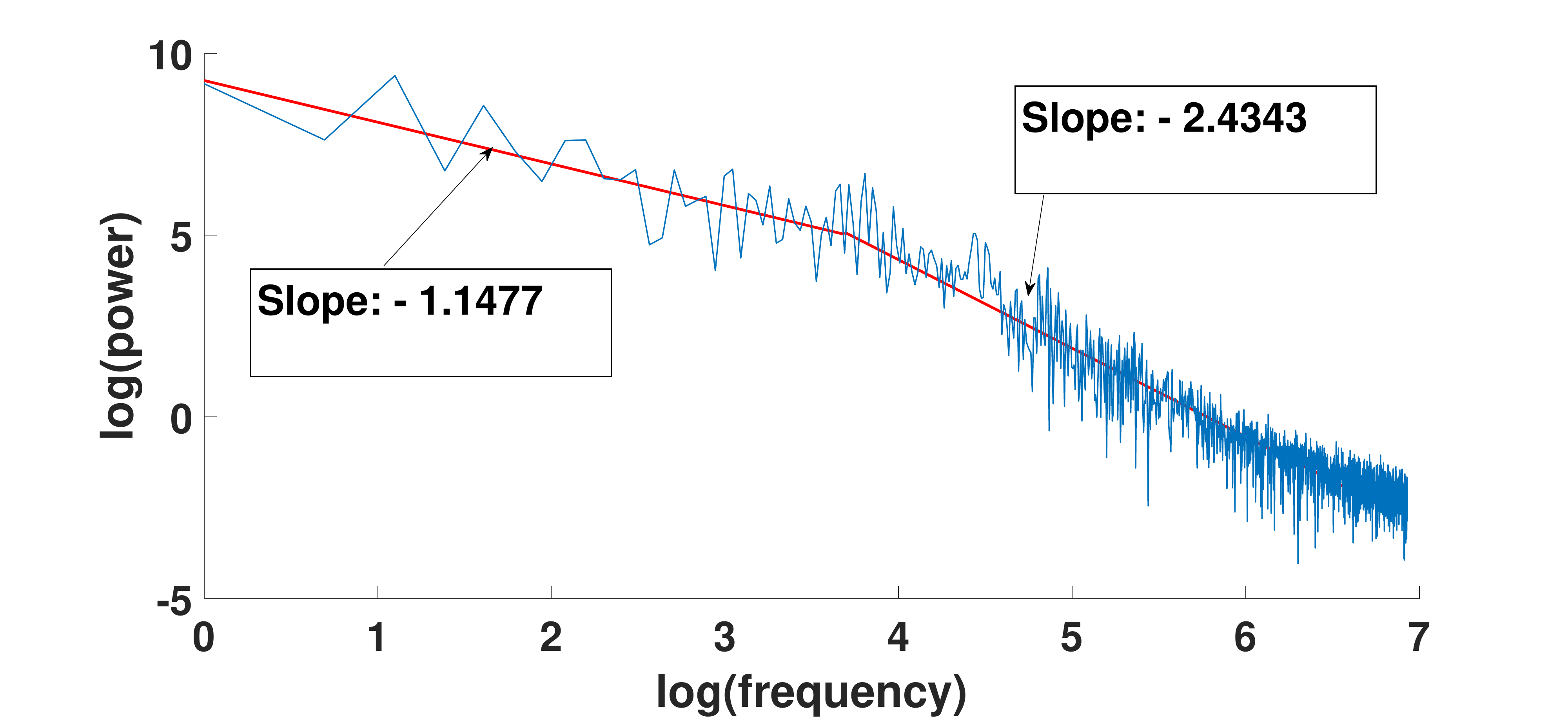}%
}
\subfloat[]{%
  \includegraphics[width=60mm,scale=0.3]{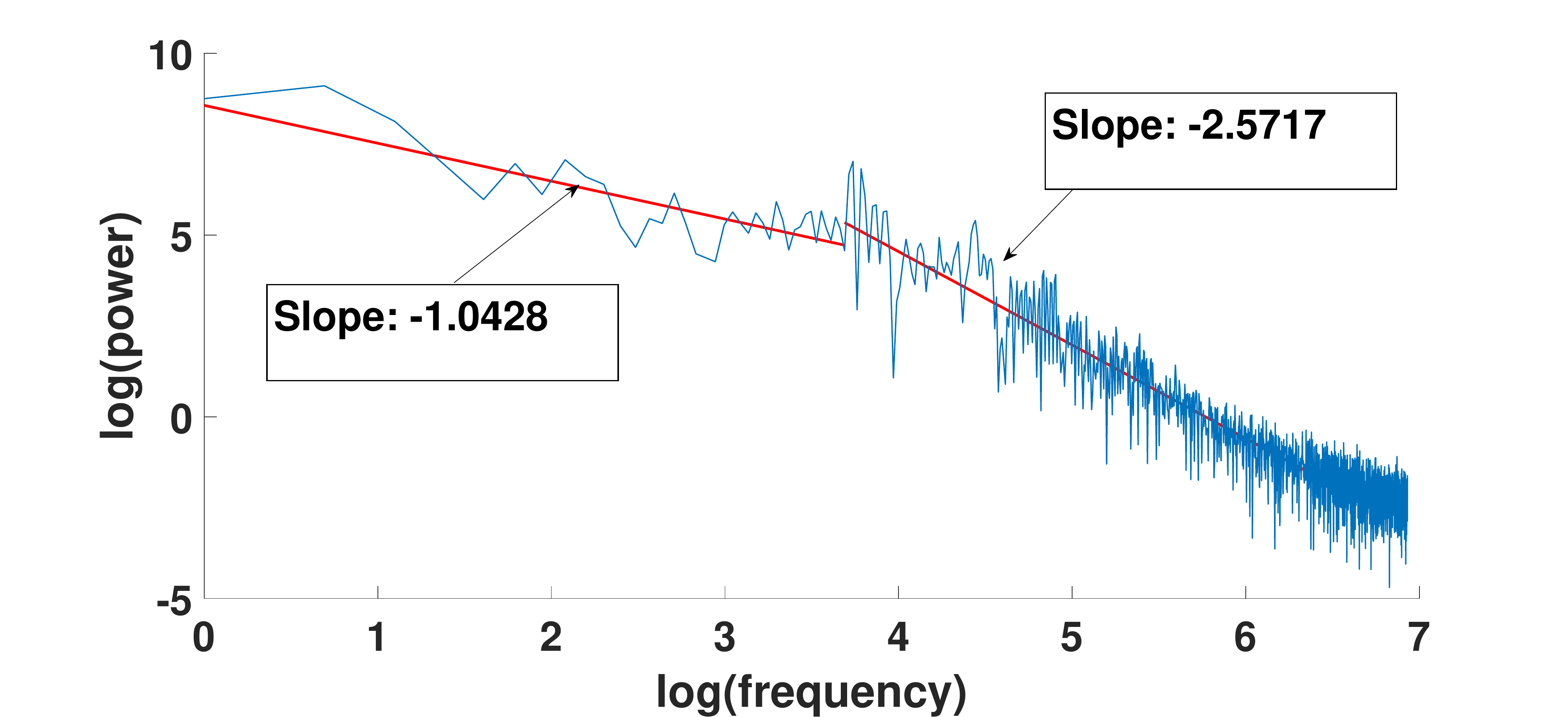}%
  }\hfill
\caption{(a),(b) are the log-log plots generated from the 50th and 350th rows of Image 1 where log(power) vs log(frequency) plots are drawn.(c),(d) is the log-log plots generated from the 50th and 350th rows of Image 4 where log(power) vs log(frequency) plots are drawn. (e),(f) are the log-log plots generated from the 50th and 350th rows of Image 12 where log(power) vs log(frequency) plots are drawn.}
\end{figure*}

\begin{table}[]
\centering
  \begin{threeparttable}
  \caption{\label{tab:table2}Comparison of Hurst Exponent values from DFA and RS analysis}
   \begin{tabular}{lllll}
 
 $I$\footnotemark[1]   &$D$\footnotemark[2] &$H$\footnotemark[3]
&$H$\footnotemark[4]  &$Remarks$\\
Image 1 &50 - 150	&0.7867 &0.7316 &Correlated\\
        &250 - 350	&0.8013	&0.8118 &Correlated\\

Image 4 &50 - 150	&0.7477	&0.7223	&Correlated\\
        &250 - 350	&0.8108	&0.6676	&Correlated\\

Image 12 &50 - 150	&0.5872	&0.6179	&Correlated\\
        &250 - 350	&0.5458	&0.5109	&Correlated\\
         \end{tabular}
    \begin{tablenotes}
      \item[1]{Image file names from Image 1 - Image 12.}
      \item[2]{Data points taken}
      \item[3]{Hurst Exponent from DFA}
      \item[4]{Hurst Exponent from RS Analysis}
    \end{tablenotes}
  \end{threeparttable}
\end{table}

\begin{table}[H]
\centering
  \begin{threeparttable}
  \caption{\label{tab:table2}Comparison of Fractal Dimension values from Hurst exponents and power-law exponent}
   \begin{tabular}{llllll}
 
 $I$\footnotemark[1]   &$D$\footnotemark[2] &$FD$\footnotemark[3]
&$FD$\footnotemark[4] &$FD$\footnotemark[5] &$FD$\footnotemark[6]  \\

Image 1 &50 - 150	&1.2133 &1.2684 &1.7587 &1.3318\\
        &250 - 350	&1.1987	&1.1812 &1.7334 &1.2599\\

Image 4 &50 - 150	&1.2523	&1.2777	&1.7598 &1.2853\\
        &250 - 350	&1.1892	&1.3324	&1.7407 &1.3355\\

Image 12 &50 - 150	&1.4182	&1.3821	&1.9261 &1.2828\\
        &250 - 350	&1.4542	&1.4891	&1.9786 &1.2141\\
        
         \end{tabular}
    \begin{tablenotes}
      \item[1]{Image file names from Image 1 - Image 12.}
      \item[2]{Data points taken}
      \item[3]{Fractal Dimension calculated from DFA}
      \item[4]{Fractal Dimension calculated from R/S analysis}
      \item[5]{Fractal Dimension calculated from power law exponent slope 1}
      \item[6]{Fractal Dimension calculated from power law exponent slope 2}
    \end{tablenotes}
  \end{threeparttable}
\end{table}

\begin{table}[h]
\centering
  \begin{threeparttable}
\caption{\label{tab:table2}Largest Lyapunov exponent values for 50th row and 350th row for respective Image 1, Image 4 and Image 12}
 \begin{tabular}{lllll}

 $I$\footnotemark[1]    &$m$\footnotemark[2] &$\tau$\footnotemark[3]
&$\lambda$\footnotemark[4]\\

Image 1 - 50th row  &10	&40 &0.2580\\
Image 1 - 350th row	&9  &30 &8.4282\\
Image 4 - 50th row	&3  &35 &2.7503\\
Image 4 - 350th row	&3  &43 &2.6087\\
Image 12 - 50th row &7	&38	&3.7984\\
Image 12 - 350th row &7 &38 &3.0345\\
         \end{tabular}
    \begin{tablenotes}
      \item[1]{File name: Image1 - Image 12}
      \item[2]{m - Embedding Dimension}
      \item[3]{$\tau$ - Time Delay}
      \item[4]{$\lambda$ - Largest Lyapunov Exponent}
    \end{tablenotes}
  \end{threeparttable}
\end{table}

\section{\label{sec:level1} Self-similarity and multifractality in spatio-temporal  patterns}
A system going through evolution due to time exposure and varying control parameter values is famously known to produce a certain behavior which is not unique to a specific system, rather shared by a large class of systems~\cite{Ref. 51}. Bak, Tang, Wiesenfeld~\cite{Ref. 52} in 1987 have discovered that dynamical systems with many interacting constituents indeed may exhibit a general characteristic behavior in which the dynamical systems organize themselves in a complex method such that no single time or length exist to control the temporal evolution of the system producing scale-invariance. Surprisingly however, the associated statistical characteristics of the complex response can be explained using a simple power law.

\subsection{\label{sec:level2}Spectral Analysis using Fourier Transform }

 Spectral analysis is a reversible technique via which a function is decomposed into a superposition of components, each with a specific temporal (or spatial) frequency.~\cite{Ref. 53} In other words, it samples or chops down a signal over a period of time (or space) and divides it into its frequency components. 
 One of the processes to study the patterns underlying in the frequency distribution is Fourier Transform, which is a linear method. With the help of Discrete Fourier Transform (DFT) technique, achieved efficiently using FFT algorithm, we have obtained Fourier Periodgram or power versus frequency plot of the available spatial data. 


The purpose of this estimation is to find the periodicities in the data, by observing peaks at the frequencies corresponding to the periodicities. The divided components are single sinusoidal oscillations at distinct frequencies having their own amplitude and phase information. For a particular dataset, if the power versus frequency plot gives two sharp peaks then that means two frequency components are dominant in the data set. 

It is well-known that a power-frequency curve of chaotic signal shows a continuous spectra over a limited range and the energy is spread over a wider bandwidth. On the other hand, power vs frequency plot of a periodic signal presents discrete spectra, where a finite number of frequencies contribute to the response. 

In our investigation, Fourier Periodgram plot shows a continues broadband spectra for a specific range which in turn suggests the data being chaotic in nature, for the set of parameters chosen for this analysis~\cite{Ref. 54, Ref. 55, Ref. 56}. Although Fourier analysis is known to be useful for measuring the frequency content of stationary or transient signals, they fail to distinguish one chaotic data from the other. As per the Fourier periodgram plots (Figure 5) received in our case, we see a broadband appearance of frequencies. In this case frequency means the number of times a pattern has been repeated over a specific data length.

\subsection{\label{sec:level2} Power spectral density analysis}

The power law can be described as a functional relationship between two quantities, where a relative change in one quantity results in a proportional relative change in the other quantity. Here, one quantity is related to power of another quantity. The main reason to bring power spectral density along with Detrended fluctuation and Rescaled range analyses explained immediately after, is to understand the multifractal nature of the data. 

Mathematically, power spectral density~\cite{Ref. 57} represented by $s(f)$ satisfies the following relation - 
\begin{equation}
 s(f) = f^{-\beta}
\end{equation}

where s(f) represents power spectral density and $\beta$ represents the power law scaling factor~\cite{Ref. 58}. In case of our data $\beta = 2H+1$ where $H$ is the Hurst exponent~\cite{Ref. 59}. This behavior produces the linear relationship when logarithms are taken of both $s(f)$ and $f$, and the straight-line on the log-log plot is often called the signature of a power law. The power law scaling factor $\beta$ can be expressed as the roughness of a spatial series (or time series) with smoother self-similar series having larger values of $\beta$.
 $0\leq \beta \leq 1$ translates to weak long range correlation whereas $1\leq \beta \leq 2$ signifies strong long range correlation. Another way to quantify the fractal nature of the data is given by fractal dimension~\cite{Ref. 60} $(D)$, that comments about the density of fractals. It is expressed by the following equation - 
\begin{equation}
\begin{split}
 D & = (3 - \gamma)/2,\hspace{.2cm}  where \hspace{.2cm}         \gamma \leq 1\\
 & = (5 - \gamma)/2,  \hspace{.2cm} where \hspace{.2cm} 1 \leq \gamma \leq 2 \\
 & = (7 - \gamma)/2, \hspace{.2cm}  where \hspace{.2cm} \gamma \geq 3 \\
\end{split}
\end{equation}
where $\gamma = \left| \beta \right|$
\\
The power law scaling factor ($\beta$) of our data in FIG. 7 shows two different values for two specific regions of a log-log plot. The Beta values for slope 1 and slope 2 of 50th row and 350th row of Image 1, Image 4 , Image 12 come out to be -1.48 and -2.33, -1.53 and -2.48, -1.48, -2.43 and -1.52, -2.33, -1.15 and -2.43, -1.04 and -2.57 with D correspondingly equals to 1.7587 and 1.3318, 1.7334 and 1.2599, 1.7598 and 1.2853 ,  1.7407 and 1.3355, 1.9261 and 1.2828, 1.9786 and 1.2141, hinting to the presence of strong range correlation.

\section{\label{sec:level1} Summary}
The central idea behind this work is calculating fractal dimension of the spatial series (analogous to time series) generated from SH equation for identifying possible presence of fractal behavior. The spatio-temporal patterns were converted into a spatial series data. The chaotic behaviour of this data has been established by estimating the positive Lyapunov exponents, phase space and power spectrum analyses. With the help of Detrended fluctuation and Rescaled range analyses, we have calculated the Hurst exponent and found the value to be greater than 0.5. This indicates that the data is long range correlated~\cite{Ref. 61}. Estimation of the Hurst exponent from both DFA and R/S analyses lead to the calculation of Fractal Dimensions that lies $\in (1,2)$. Recently, both long range correlation and chaotic behavior~\cite{Ref. 62} has been observed by researchers while analysing the genomes of SARS-CoV (2002) and SARS-CoV-2 (2019), responsible for the global pandemic(COVID-19). Our method can aid in analysing similar nonlinear time series and spatial series.


Further, we have carried out an estimation of Fourier power law analysis which confirms that the exponent value ($\beta$) lies between $-1.04$ to $-2.57$ for the entire data set. We have also calculated the fractal dimension from the power law exponent values and found them comparable to the data generated from DFA and R/S analysis. The power law (log power-log frequency plot) scaling factor highlights the presence of more than one exponents. 
Whenever the fractal dimension, as a single exponent is not enough to describe the complexity of the system, multifractality can generalize the complexity by using several exponent, capturing its dynamical essence. As seen in our case, different scaling exponents were essential for different segments of the same spatial series, indicating occurrence of the scaling behaviour and characterizing multifractality in our data set. In the patterns, this may indicate interlinked fractal subsets in the spatial series~\cite{Ref. 63} and hence, our work can be used for analysing and studying systems with interlinked fractal subsets and multifractality behavior.


\section*{\label{sec:level1} Acknowledgments}D.B., A.K.J, A.N.S.I, and M.S.J would like to acknowledge the director, SINP, for his constant encouragement for the project. A.N.S.I would like to thank DST-SERB for their ardent support.  


\end{document}